\documentclass[leqno,11pt]{article}
\usepackage{amsfonts,latexsym}
\usepackage{amsmath}
\usepackage{amscd}
\usepackage{float,amsmath,amssymb,mathrsfs,bm,multirow,graphics}
\usepackage[dvips]{graphicx}

\newcommand{\nequation}{\setcounter{equation}{0}}
\renewcommand{\theequation}{\mbox{\arabic{section}.\arabic{equation}}}
\newcommand{\R}{{\Bbb R}}

\newcommand{\C}{{\Bbb C}}

\newcommand{\Q}{{\Bbb Q}}
\newcommand{\proofbegin}{\noindent{\it Proof.\,\,}}
\newcommand{\proofend}{\hfill$\Box$\bigskip}
\newtheorem{theorem}{Theorem}[section]
\newtheorem{proposition}[theorem]{Proposition}
\newtheorem{lemma}[theorem]{Lemma}
\newtheorem{definition}[theorem]{Definition}
\newtheorem{assumption}[theorem]{Assumption}
\newtheorem{remark}[theorem]{Remark}

\newtheorem{figuretext}[theorem]{Figure}

\input epsf
\title{\sc An Integrable Generalization of the Nonlinear Schr\"odinger Equation on the Half-Line and Solitons}
\author{J. Lenells and A. S. Fokas}
\date{{\small Department of Applied Mathematics and Theoretical Physics, University of Cambridge, Cambridge CB3 0WA, United Kingdom}}

\begin{document}
\maketitle
\begin{abstract} 
\noindent
We analyze initial-boundary value problems for an integrable generalization of the nonlinear Schr\"odinger equation formulated on the half-line. In particular, we investigate the so-called linearizable boundary conditions, which in this case are of Robin type. Furthermore, we use a particular solution to verify explicitly all the steps needed for the solution of a well-posed problem.
 \end{abstract}



\section{Introduction}\label{introsec}\nequation
A novel integrable generalization of the KdV equation, the so-called Camassa-Holm equation, was derived from mathematical and physical considerations in \cite{F-F} and \cite{C-H} respectively (see also \cite{F-Liu}). The following analogous integrable nonlinear equation associated with the nonlinear Schr\"odinger (NLS) equation was derived in \cite{F}: 
\begin{subequations}
\begin{equation}\label{GNLS} 
  iu_t - \nu u_{tx} + \gamma u_{xx} + \sigma |u|^2(u  + i \nu u_x) = 0, \qquad \sigma = \pm 1,
\end{equation}
where $u(x,t)$ is a complex-valued function and $\gamma$ and $\nu$ are real constants. The initial value problem of equation (\ref{GNLS}) was analyzed in \cite{L-F}.

Replacing $u(x,t)$ by $u(-x,t)$ in (\ref{GNLS}) if necessary, we can assume that $\gamma/\nu > 0$.
Then the gauge transformation $u \to \sqrt{\gamma/\nu^3} \,\mathrm{exp}(ix/\nu) u$
transforms (\ref{GNLS}) into
$$u_{tx} + \frac{\gamma}{\nu^3} u - \frac{2i\gamma}{\nu^2}u_x - \frac{\gamma}{\nu} u_{xx} - \frac{i \gamma}{\nu^3}\sigma |u|^2u_x = 0.$$
For simplicity we set $\gamma = \nu = -\sigma = 1$ in the rest of this paper and consider the equation
\begin{equation}\label{GNLS2}
  u_{tx} + u - 2iu_x - u_{xx} + i |u|^2u_x = 0.
\end{equation}
\end{subequations}

In this paper: (a) We study equation (\ref{GNLS2}) on the half-line by employing the formalism of \cite{F1997} (see also \cite{F2002, F-I-S}). (b) We solve explicitly some concrete initial-boundary value (IBV) problems corresponding to the so-called linearizable boundary conditions.

Regarding (a) we note that the formalism of \cite{F1997} expresses the solution of an IBV problem on the half-line in terms of the solution of a matrix-valued Riemann-Hilbert (RH) problem, which has jump matrices with an explicit exponential $(x,t)$ dependence and which are defined in terms of the so-called spectral functions denoted by $\{a(\zeta), b(\zeta), A(\zeta), B(\zeta)\}$. The functions $a(\zeta)$ and $b(\zeta)$ can be defined in terms of the initial data $u_0(x)$ via a linear Volterra integral equation, whereas the functions $A(\zeta)$ and $B(\zeta)$ can be defined in terms of the boundary values $u(0,t)$ and $u_x(0,t)$ also via a linear Volterra integral equation. However, for a well-posed problem only one of the functions $u(0,t)$ and $u_x(0,t)$ (or their combination) is prescribed as boundary conditions. Thus, in order to compute $A(\zeta)$ and $B(\zeta)$, one must first characterize the unknown boundary values in terms of the given boundary data and of $u_0(x)$. Unfortunately, the solution of this problem, which makes crucial use of the so-called global relation, involves a nonlinear Volterra integral equation \cite{B-F-S, F2005}. In spite of this difficulty, the above formulation does yield essential information about the solution, such as its long time behavior \cite{F-I1992, F-I1994, F-I1996}.

Regarding (b) we note that there exists a particular class of boundary conditions, called linearizable, for which it is possible to bypass the difficulty of computing the unknown boundary values. In this case, it is possible to express directly $A(\zeta)$ and $B(\zeta)$ in terms of $a(\zeta)$ and $b(\zeta)$ and the given boundary data. Thus the relevant formalism, which was developed in \cite{F2002}, is as effective as the usual inverse scattering transform formalism.

Expressions relating $A(\zeta)$ and $B(\zeta)$ to $a(\zeta)$ and $b(\zeta)$ for several linearizable boundary value problems have been presented in several papers, see for example \cite{F2002, F2004, F-I-S}. However, to the best of our knowledge, the relevant formalism has never been implemented explicitly. Here we will present examples of an explicit implementation. In particular, it will be shown in section \ref{linearizablesec} that the following class of boundary conditions is linearizable:
\begin{equation}
  u_x(0, t) = u(0, t) e^{i\alpha}, \qquad \alpha \in \R.
\end{equation}
It was shown in \cite{L-F} that on the full line there exists a four-parameter family of one-soliton solutions. Let $u_0(x) = u(x,0)$, $x > 0$, be the restriction of a one-soliton solution to $x > 0$ at $t = 0$. Then, $u(x,t)$, $x > 0$, provides the solution of the IBV problem with initial data $u_0(x)$ and boundary values $u(0,t)$ and $u_x(0,t)$. It turns out that a three-parameter subfamily of the one-soliton solutions, which will be denoted by $u^s(x,t)$, satisfy a linearizable boundary condition. In particular, the following initial and boundary conditions define a linearizable IBV problem:
\begin{equation}
  u_0^s(x) = F(x; \gamma, x_0, \Sigma_0), \quad x > 0, \qquad u_x^s(0,t) = e^{i \alpha} u^s(0,t), \quad t > 0,
\end{equation}
where the function $F$ defined by
\begin{equation}
  F (x; \gamma, x_0, \Sigma_0) = -\frac{2 i \sqrt{2} e^{(x - x_0) \sin{\gamma} - i (\gamma +2
   \Sigma_0 -x \cos{\gamma})} \sin{\gamma}}{e^{i \gamma}+e^{2 (x - x_0) \sin{\gamma}}},
\end{equation} 
depends on the three parameters $\gamma \in (0, \pi)$, $x_0 \in \R$, $\Sigma_0 \in \R$, and $e^{i \alpha}$ is defined by
\begin{equation}
  e^{i \alpha} = i \cos \left(\frac{1}{2} (\gamma + 2 i x_0 \sin{\gamma} )\right) \sec \left(\frac{1}{2} (\gamma - 2 i x_0 \sin{\gamma} )\right).
\end{equation}
In section \ref{solitonsec} we will consider the spectral analysis of the Lax pair equations associated with $u^s(x,t)$. Using the standard inverse scattering method it is straightforward to determine the eigenfunctions and scattering data on the full line corresponding to $u^s(x,t)$. By relating the Lax pair of the full-line problem to the one of the half-line problem, we can obtain the eigenfunction $\mu_3(x,t,\zeta)$ needed for the half-line formulation ($\mu_3$ is defined by integrating from $x = \infty$, just like one of the eigenfunctions in the spectral analysis on the line). Then, by evaluating $\mu_3(x,t,\zeta)$ at $x = t = 0$, we can find explicit expressions for the spectral functions $a(\zeta)$ and $b(\zeta)$. In the problem on the line the zeros of the spectral function $a(\zeta)$ are directly linked to solitons, and for a pure soliton solution $b(\zeta)$ vanishes identically. This is not the case for the half-line problem. Nevertheless, we find that as the location of the center-of-mass of the initial data $u^s(x,0)$ approaches infinity, the zeros of $a(\zeta)$ approach the corresponding zeros for the problem on the line and $b(\zeta) \to 0$. This is consistent with the fact that in this limit the boundary should have no effect on the formulation of the problem. 

In sections \ref{exampledirectsec} and \ref{examplelinearizablesec} we analyze the particular case of 
$$\gamma = \frac{\pi}{2}, \qquad x_0 = 0, \qquad \Sigma_0 = 0.$$
The particular solution obtained for these parameter values, which will be denoted by $u^p(x,t)$, satisfies the following initial and boundary conditions:
\begin{equation*}
  u_0^p(x) = -\frac{2 \sqrt{2} e^x}{i+e^{2 x}}, \quad x > 0,	 \qquad  u_x^p(0,t) =  i u^p(0,t), \quad t > 0.
\end{equation*}

In section \ref{exampledirectsec} we compute explicitly all eigenfunctions $\mu_j(x,t,\zeta)$, $j = 1, 2, 3$, needed for the formulation of the basic RH problem, as well as the spectral functions $a(\zeta)$, $b(\zeta)$, $A(\zeta)$, $B(\zeta)$. We then state the RH problem for $M$ and verify by direct computation that $M$ satisfies the correct jump and residue conditions. In section \ref{examplelinearizablesec}, we apply the general formalism developed for linearizable boundary conditions and show that it reconstructs the solution 
\begin{equation}\label{onesolitonintro}  
  u^p(x,t) = -\frac{2 \sqrt{2} e^{2 i t+x}}{i+e^{2 x}}.
\end{equation}
In the case of the half-line, it is not the zeros of $a(\zeta)$ that are linked to solitons but the zeros of another function, denoted by $d(\zeta)$, whose definition involves all the spectral functions $a(\zeta)$, $b(\zeta)$, $A(\zeta)$, and $B(\zeta)$ (see equation (\ref{ddef}) below). In the linearizable case, $d(\zeta)$ can be effectively replaced by another function $\mathcal{D}(\zeta)$ whose definition involves only $a(\zeta)$ and $b(\zeta)$. We verify for our example that in the long time asymptotics, the zeros of $\mathcal{D}(\zeta)$ indeed correspond to the one-soliton (\ref{onesolitonintro}).

\section{Spectral analysis}\label{spectralsec}\nequation

\subsection{A Lax pair}
Let
\begin{equation}\label{etadef}
 U(x,t) =\begin{pmatrix} 0	& u(x,t)	\\
v(x,t)	&	0 \end{pmatrix}, \qquad \sigma_3 = \begin{pmatrix} 1 & 0 \\ 0 & -1 \end{pmatrix}, \qquad \eta = \zeta - \frac{1}{2\zeta}, \qquad v = \bar{u}.
\end{equation}
Equation (\ref{GNLS2}) is the condition of compatibility of
\begin{equation}\label{psilax}
\begin{cases}
	& \psi_x + i\zeta^2 \sigma_3 \psi = \zeta U_x \psi,	\\
	& \psi_t + i\eta^2 \sigma_3 \psi = \left(\zeta U_x - \frac{i}{2}\sigma_3U^2 +\frac{i}{2\zeta}\sigma_3U\right)\psi,
\end{cases}
\end{equation}
where $\psi(x,t,\zeta)$ is a $2\times2$ matrix valued function and $\zeta \in \C$ is a spectral parameter.
Starting with this Lax pair and following steps similar to the ones used in \cite{L-F}  (where now we integrate with respect to $x$ starting from $x=0$ instead of $x = -\infty$) we find that in order to have a function satisfying, within its region of boundedness,
\begin{equation}\label{muasymptotic}  
  \mu = I + O\left(\frac{1}{\zeta}\right), \qquad \zeta \to \infty,
\end{equation}
we make the substitution
\begin{equation}\label{psimurelation}
  \psi(x,t,\zeta) = e^{i\int^{(x,t)}_{(0, 0)} \Delta \sigma_3}\mu(x,t,\zeta) e^{-i\int_{(0,0)}^{(\infty, 0)} \Delta \sigma_3}e^{-i(\zeta^2x + \eta^2t)\sigma_3},
\end{equation}
where $\Delta$ is the closed real-valued one-form
\begin{equation}\label{Deltadef}  
  \Delta(x,t) = \frac{1}{2}u_xv_x dx + \frac{1}{2}(u_xv_x - uv)dt.
\end{equation}
The function $\mu$ satisfies the following Lax pair:
\begin{equation}\label{mulax}  
\begin{cases}
	& \mu_x + i\zeta^2 [\sigma_3, \mu] = V_1\mu, \\
	& \mu_t + i\eta^2 [\sigma_3, \mu] = V_2\mu,
\end{cases}
\end{equation}
where the matrices $V_1$ and $V_2$ are defined by
\begin{align}\label{V1explicit}
V_1 =& \begin{pmatrix} -\frac{i}{2} u_xv_x	&	\zeta u_x e^{-2i\int^{(x,t)}_{(0, 0)} \Delta} \\
\zeta v_x e^{2i\int^{(x,t)}_{(0, 0)} \Delta}	&	\frac{i}{2} u_xv_x \end{pmatrix},
	\\	\label{V2explicit}
V_2 =& \begin{pmatrix} -\frac{i}{2}u_xv_x 		&	\left(\zeta u_x + \frac{i}{2\zeta} u\right)e^{-2i\int^{(x,t)}_{(0, 0)} \Delta}	\\
\left(\zeta v_x - \frac{i}{2\zeta} v\right)e^{2i\int^{(x,t)}_{(0, 0)} \Delta}	&	\frac{i}{2}u_xv_x	\end{pmatrix}.
\end{align}
Equation (\ref{mulax}) can be written as the following single equation:
\begin{equation}\label{laxdiffform}  
  d\left(e^{i(\zeta^2 x + \eta^2 t)\hat{\sigma}_3} \mu(x,t,\zeta) \right) = W(x,t,\zeta),
\end{equation}
where
$$W(x,t,\zeta) = e^{i(\zeta^2 x + \eta^2 t)\hat{\sigma}_3} \left(V_1(x,t,\zeta) dx + V_2(x,t,\zeta) dt\right) \mu(x,t,\zeta)$$
and $\hat{\sigma}_3$ acts on a $2\times 2$ matrix $A$ by $\hat{\sigma}_3A = [\sigma_3, A]$.

\subsection{Bounded and analytic eigenfunctions}
\begin{figure}
\begin{center}
    \includegraphics[width=.3\textwidth]{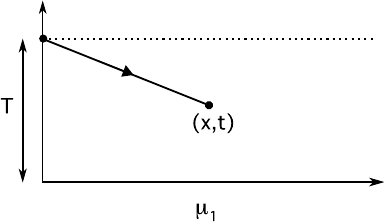} \quad
    \includegraphics[width=.3\textwidth]{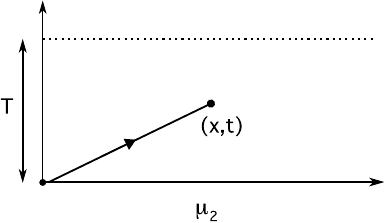} \quad
    \includegraphics[width=.3\textwidth]{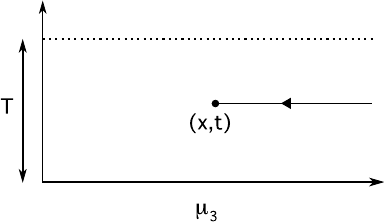} \\
     \begin{figuretext}\label{mucontours.pdf}
       The contours of integration used for the definition of the solutions $\mu_1$, $\mu_2$ and $\mu_3$ of (\ref{mujdef}).
     \end{figuretext}
     \end{center}
\end{figure}

Let equation (\ref{laxdiffform}) be valid for 
$$0 < x < \infty, \qquad 0 < t< T,$$
where $T \leq \infty$ is a given positive constant; unless otherwise stated, we suppose that $T < \infty$.
Assuming that the function $u(x,t)$ has sufficient smoothness and decay, we define three solutions $\mu_j$, $j = 1,2,3$, of (\ref{laxdiffform}) by
\begin{equation}\label{mujdef}  
  \mu_j(x,t,\zeta) = I + \int_{(x_j, t_j)}^{(x,t)} e^{-i(\zeta^2 x + \eta^2 t)\hat{\sigma}_3}W(x',t',\zeta),
\end{equation}
where $(x_1, t_1) = (0, T)$, $(x_2, t_2) = (0, 0)$, and $(x_3, t_3) = (\infty, t)$. Since the one-form $W$ is exact, the integral on the right-hand side of equation (\ref{mujdef}) is independent of the path of integration. We choose the particular contours shown in Figure \ref{mucontours.pdf}. This choice implies the following inequalities on the contours:
\begin{align*}
(x_1, t_1) \to (x,t): x' - x \leq 0,& \qquad t' - t \geq 0,
	\\
(x_2, t_2) \to (x,t): x' - x \leq 0,& \qquad t' - t \leq 0,
	\\
(x_3, t_3) \to (x,t): x' - x \geq 0.&
\end{align*}
The second column of the matrix equation (\ref{mujdef}) involves $\exp[2i(\zeta^2(x' - x) + \eta^2(t' - t))]$. 
Using the above inequalities it follows that this exponential is bounded in the following regions of the complex $\zeta$-plane:
\begin{align*}
(x_1, t_1) \to (x,t): \{\text{Im}\, \zeta^2 \leq 0\} &\cap \{\text{Im}\, \eta^2 \geq 0\},
	\\
(x_2, t_2) \to (x,t): \{\text{Im}\, \zeta^2 \leq 0\} &\cap \{\text{Im}\, \eta^2 \leq 0\},
	\\
(x_3, t_3) \to (x,t): \{\text{Im}\, \zeta^2 \geq 0\}&.
\end{align*}
Define the following sets (see Figure \ref{GNLSRH2.pdf}):
\begin{align} \nonumber
&D_1 = \{\zeta \in \C| \arg \zeta \in (0, \pi/2) \cup (\pi, 3\pi/2) \text{  and  } |\zeta| > 1/\sqrt{2}\},
	\\ \label{D1234def}
&D_2 = \{\zeta \in \C| \arg \zeta \in (0, \pi/2) \cup (\pi, 3\pi/2) \text{  and  } |\zeta| < 1/\sqrt{2}\},
	\\ \nonumber
&D_3 = \{\zeta \in \C| \arg \zeta \in (\pi/2, \pi) \cup (3\pi/2, 2\pi) \text{  and  } |\zeta| < 1/\sqrt{2}\},
	\\ \nonumber
&D_4 = \{\zeta \in \C| \arg \zeta \in (\pi/2, \pi) \cup (3\pi/2, 2\pi) \text{  and  } |\zeta| > 1/\sqrt{2}\}.
\end{align}
Since
$$\eta = \zeta - \frac{1}{2\zeta} = \left(1 - \frac{1}{2|\zeta|^2}\right)\mathrm{Re}\,\zeta + i\left(1 + \frac{1}{2|\zeta|^2}\right)\mathrm{Im}\,\zeta,$$
we deduce that the second column vectors of $\mu_1$, $\mu_2$, and $\mu_3$ are bounded and analytic for $\zeta \in \C$ provided that $\zeta$ belongs to $D_3$, $D_4$, and $D_1 \cup D_2$, respectively.
We will denote these vectors with superscripts $(3)$, $(4)$, and $(12)$ to indicate the domains of their boundedness.
Similar considerations are valid for the first column vectors. Thus,
\begin{equation}\label{mu123boundedness}
\mu_1 = (\mu_1^{(2)}, \mu_1^{(3)}), \qquad 
\mu_2 = (\mu_2^{(1)}, \mu_2^{(4)}), \qquad 
\mu_3 = (\mu_3^{(34)}, \mu_3^{(12)}).
\end{equation}

Note that the eigenfunctions $\mu_1$ and $\mu_2$ are defined and analytic in the whole complex $\zeta$-plane except at $\zeta = \infty$ and at $\zeta = 0$, where they, in general, have essential singularities. By (\ref{muasymptotic}), the column vectors of the $\mu_j$'s approach the corresponding column vectors of the identity matrix as $\zeta \to \infty$ within their regions of boundedness, i.e.
$$\left(\mu_2^{(1)}(x, t, \zeta), \mu_3^{(12)}(x,t,\zeta)\right) = I+ O\left(\frac{1}{\zeta}\right), \qquad \zeta \to \infty, \quad \zeta \in D_1,$$
$$\left(\mu_3^{(34)}(x, t, \zeta), \mu_2^{(4)}(x,t,\zeta)\right) = I+ O\left(\frac{1}{\zeta}\right), \qquad \zeta \to \infty, \quad \zeta \in D_4.$$
The functions $\{\mu_j\}_{j = 1}^3$ are the fundamental eigenfunctions needed for the formulation of a Riemann-Hilbert problem in the complex $\zeta$-plane. Indeed, for each region $D_j$, $j=1, \dots, 4$, of the complex $\zeta$-plane there exist two column vectors which are bounded and analytic in $D_j$ and which have bounded and continuous extensions to $\bar{D}_j$.  For example, in $D_1$ these two vectors are $\mu_2^{(1)}$ and $\mu_3^{(12)}$.

Observe that the Lax pair equations (\ref{psilax}) are of `standard form' near the singularities at $\zeta = \infty$ and $\zeta = 0$. As $\zeta \to \infty$, the highest-order terms of $O(\zeta^2)$ of the $x$ and $t$ parts involve the {\it diagonal} matrix $i\zeta^2 \sigma_3$, whereas the subleading terms of $O(\zeta)$ involve the {\it off-diagonal} matrix $\zeta U_x$. Similarly, as $\zeta \to 0$, the highest-order term of $O(1/\zeta^2)$ of the $t$-part involves the {\it diagonal} matrix $\frac{i}{4\zeta^2} \sigma_3$, whereas the subleading term of $O(1/\zeta)$ involves the {\it off-diagonal} matrix $\frac{i}{2\zeta}\sigma_3U$. In the integral equation (\ref{mujdef}) the exponent contains the highest-order terms, while the one-form $W$ contains the subleading terms. In particular, the boundedness properties (\ref{mu123boundedness}) of the eigenfunctions $\{\mu_j\}_1^3$ are valid also for $\zeta$ near $\zeta = \infty$ and $\zeta = 0$. The behavior of the $\mu_j$'s as $\zeta \to 0$ is studied in detail in Appendix \ref{appendixA}.

\subsection{Spectral functions}
In order to derive a Riemann-Hilbert problem, we must compute the `jumps' of these vectors across the boundaries of the domains $\{D_j\}_{j = 1}^4$.
It turns out that the relevant jump matrices can be uniquely defined in terms of two $2\times 2$-matrix valued spectral functions $s(\zeta)$ and $S(\zeta)$ defined as follows. 
Any two solutions $\mu$ and $\tilde{\mu}$ of (\ref{laxdiffform}) are related by an equation of the form
\begin{equation}\label{mutildemu}  
  \mu(x,t,\zeta) = \tilde{\mu}(x,t,\zeta)  e^{-i(\zeta^2 x + \eta^2 t)\hat{\sigma}_3} C_0(\zeta),
\end{equation}
where $C_0(\zeta)$ is a $2\times 2$ matrix independent of $x$ and $t$. Indeed, let $\psi$ and $\tilde{\psi}$ be the solutions of equation (\ref{psilax}) corresponding to $\mu$ and $\tilde{\mu}$ according to (\ref{psimurelation}).
Then, since the first and second column of a solution of (\ref{psilax}) satisfy the same equation, there exists a $2 \times 2$ matrix $C_1(\zeta)$ independent of $x$ and $t$ such that
\begin{equation}\label{psirelations}  
  \psi(x,t, \zeta) = \tilde{\psi}(x,t, \zeta) C_1(\zeta).
\end{equation}
Hence (\ref{mutildemu}) is valid with $C_0(\zeta) = e^{-i\int^{(\infty, 0)}_{(0, 0)} \Delta \hat{\sigma}_3} C_1(\zeta)$. 

We define $s(\zeta)$ and $S(\zeta)$ by the relations
\begin{align}
\label{seq} 
  \mu_3(x,t,\zeta) &= \mu_2(x,t,\zeta)e^{-i(\zeta^2 x + \eta^2 t)\hat{\sigma}_3} s(\zeta),
		\\
  \label{Seq} 
  \mu_1(x,t,\zeta) &= \mu_2(x,t,\zeta)  e^{-i(\zeta^2 x + \eta^2 t)\hat{\sigma}_3} S(\zeta).
\end{align}

Evaluation of (\ref{seq}) and (\ref{Seq}) at $(x,t) = (0,0)$ and $(x,t) = (0,T)$ gives the following expressions:
\begin{equation}\label{Ssdef}   
   s(\zeta) = \mu_3(0,0, \zeta), \qquad S(\zeta) = \mu_1(0,0, \zeta) = \left(e^{i\eta^2T\hat{\sigma}_3}\mu_2(0,T,\zeta)\right)^{-1}.
\end{equation}
Hence, the functions $s(\zeta)$ and $S(\zeta)$ can be obtained from the evaluations at $x = 0$ and at $t = T$ of the functions $\mu_3(x,0, \zeta)$ and $\mu_2(0,t,\zeta)$, which satisfy the linear integral equations
\begin{equation}\label{mu3x0equation}
  \mu_3(x,0,\zeta) = I + \int_{\infty}^{x} e^{i\zeta^2 (x' -x) \hat{\sigma}_3}(V_1\mu_3)(x',0,\zeta) dx',
\end{equation}
and
\begin{equation}\label{mu20tequation}
   \mu_2(0,t,\zeta) = I + \int_{0}^{t} e^{i\eta^2 (t' -t) \hat{\sigma}_3}(V_2\mu_2)(0,t',\zeta) dt'.
\end{equation}
By evaluating (\ref{V1explicit}) and (\ref{V2explicit}) at $t = 0$ and $x = 0$, respectively, we find that
\begin{equation}\label{V1initial}
V_1(x,0, \zeta) = \begin{pmatrix} -\frac{i}{2} |u_{0x}|^2	&	\zeta u_{0x} e^{-i \int^x_{0} |u_{0x}|^2 dx'} \\
\zeta \bar{u}_{0x} e^ {i \int^x_{0} |u_{0x}|^2 dx'}	&	\frac{i}{2} |u_ {0x}|^2 \end{pmatrix}
\end{equation}
and
\begin{align}\label{V2boundary}
 V_2(0,t,\zeta) =
 \begin{pmatrix} -\frac{i}{2}|g_1|^2 		&	\left(\zeta g_1 + \frac{i}{2\zeta} g_0\right)e^{-i\int^{t}_{0} (|g_1|^2 - |g_0|^2)dt'}	\\
 \left(\zeta \bar{g}_1 - \frac{i}{2\zeta} \bar{g}_0\right)e^{i\int^{t}_{0} (|g_1|^2 - |g_0|^2)dt'}	&	\frac{i}{2}|g_1|^2	\end{pmatrix},
\end{align}
where $u_0(x) = u(x,0)$, $g_0(t) = u(0,t)$, and $g_1(t) = u_x(0,t)$ are the initial and boundary values of $u(x,t)$.
The expressions for $V_1(x,0, \zeta)$ and $V_2(0,t,\zeta)$ contain only $u_0(x)$ and $\{g_0(t), g_1(t)\}$, respectively. Therefore, the integral equation (\ref{mu3x0equation}) determining $s(\zeta)$ is defined in terms of the initial data $u_0(x)$, and the integral equation (\ref{mu20tequation}) determining $S(\zeta)$ is defined in terms of the boundary values $g_0(t)$ and $g_1(t)$.

\subsection{Symmetries}
\begin{proposition}
Let $\{\mu_j\}_{j = 1}^3$ be defined by equation (\ref{mujdef}). Then $\mu(x,t,\zeta) = \mu_j(x,t,\zeta)$, $j= 1,2,3$, satisfies the following symmetry relations:
\begin{align}\label{musymmetries1}
  \mu_{11}(x,t,\zeta) = \overline{\mu_{22}(x,t,\bar{\zeta})}, \qquad \mu_{21}(x,t,\zeta) = \overline{\mu_{12}(x,t, \bar{\zeta})},
	\\ 
	 \label{musymmetries2}
\begin{cases}  
\mu_{12}(x, t, -\zeta) = -\mu_{12}(x, t, \zeta), \qquad \mu_{11}(x, t, -\zeta) = \mu_{11}(x, t,\zeta),
  		\\
    \mu_{21}(x, t, -\zeta) = -\mu_{21}(x, t, \zeta), \qquad \mu_{22}(x, t, -\zeta) = \mu_{22}(x, t, \zeta).
    \end{cases}
\end{align}
\end{proposition}
\proofbegin
For a $2\times 2$ matrix $A$, we define the $2\times 2$ matrices $TA$ and $PA$ by
$$TA = \begin{pmatrix} 
      \bar{a}_{22} & \bar{a}_{21} \\
      \bar{a}_{12} & \bar{a}_{11} \\
   \end{pmatrix},
   \quad PA = \begin{pmatrix} 
      a_{11} & -a_{12} \\
      -a_{21} & a_{22} \\
   \end{pmatrix}
\quad \text{where} \quad
A = \begin{pmatrix} 
      a_{11} & a_{12} \\
      a_{21} & a_{22} \\
   \end{pmatrix}.$$
Equations (\ref{musymmetries1}) and (\ref{musymmetries2}) are a consequence of the symmetries $(TV_j)( \bar{\zeta}) = V_j(\zeta)$ and $(PV_j)(-\zeta) = V_j(\zeta)$, respectively, valid for $j = 1,2$.
\proofend

\subsection{The functions $s(\zeta)$ and $S(\zeta)$}\label{abABsubsec}
If $\psi(x,t,\zeta)$ satisfies (\ref{psilax}), it follows that $\det \psi$ is independent of $x$ and $t$. Hence the determinant of $\mu$ (which is related to $\psi$ via (\ref{psimurelation})) is also independent of $x$ and $t$. In particular, for $\mu_j$, $j = 1,2,3$, the evaluation of $\det \mu_j$ at $(x_j, t_j)$ shows that
$$\det \mu_j = 1, \quad j= 1,2,3.$$
In particular,
$$\det s(\zeta) = \det S(\zeta) = 1.$$

From (\ref{musymmetries1}) it follows that
$$s_{11}(\zeta) = \overline{s_{22}(\bar{\zeta})}, \quad s_{21}(\zeta) = \overline{s_{12}(\bar{\zeta})}, \quad S_{11}(\zeta) = \overline{S_{22}(\bar{\zeta})}, \quad S_{21}(\zeta) = \overline{S_{12}(\bar{\zeta})}.$$
These symmetry relations justify the following notations for $s$ and $S$:
\begin{equation}\label{sSandabAB}
s(\zeta) = \begin{pmatrix} \overline{a(\bar{\zeta})} & b(\zeta) \\
 \overline{b(\bar{\zeta})} 	&	a(\zeta) \end{pmatrix}, \qquad 
S(\zeta) = \begin{pmatrix} \overline{A(\bar{\zeta})} & B(\zeta) \\
 \overline{B(\bar{\zeta})} 	&	A(\zeta) \end{pmatrix}.
 \end{equation}
The symmetry relations (\ref{musymmetries2}) imply that $a(\zeta)$ and $A(\zeta)$ are even functions of $\zeta$, whereas $b(\zeta)$ and $B(\zeta)$ are odd functions of $\zeta$, that is,
\begin{equation}\label{abABevenoddsymmetries}  
  a(-\zeta) = a(\zeta), \quad b(-\zeta) = -b(\zeta), \quad A(-\zeta) = A(\zeta), \quad B(-\zeta) = -B(\zeta).
\end{equation}

The definitions of $\mu_j(0,t,\zeta)$, $j = 1,2$, and of $\mu_2(x,0,\zeta)$ imply that these functions have larger domains of boundedness, namely:
\begin{equation}\label{mu1largerdomain}
  \mu_1(0,t,\zeta) = \left(\mu_1^{(24)}(0,t,\zeta), \mu_1^{(13)}(0,t,\zeta)\right),
\end{equation}
$$\mu_2(0,t,\zeta) = \left(\mu_2^{(13)}(0,t,\zeta), \mu_2^{(24)}(0,t,\zeta)\right),$$
$$\mu_2(x,0,\zeta) = \left(\mu_2^{(12)}(x,0,\zeta), \mu_2^{(34)}(x,0,\zeta)\right).$$
The definitions of $s(\zeta)$ and $S(\zeta)$ imply
\begin{equation}\label{abABdef}
\begin{pmatrix}b(\zeta) \\
a(\zeta) \end{pmatrix} = \mu_3^{(12)}(0,0,\zeta), \qquad \begin{pmatrix} -e^{-2i\eta^2 T} B(\zeta) \\
\overline{A(\bar{\zeta})} \end{pmatrix} = \mu_2^{(24)}(0, T, \zeta).
\end{equation}
Let us summarize the properties of the spectral functions.
\begin{itemize}
\item  $a(\zeta)$ and $b(\zeta)$ are continuous and bounded for $\zeta \in \bar{D}_1 \cup \bar{D}_2$ and analytic in $D_1 \cup D_2$.
\item $a(\zeta)\overline{a(\bar{\zeta})} -  b(\zeta)\overline{b(\bar{\zeta})} = 1, \qquad \zeta \in \bar{D}_1 \cup \bar{D}_2$.
\item $a(\zeta) = 1 + O\left(\frac{1}{\zeta}\right), \qquad b(\zeta) = O\left(\frac{1}{\zeta}\right), \qquad \zeta \to \infty, \quad \zeta \in D_1 \cup D_2.$
\item $A(\zeta)$ and $B(\zeta)$ are continuous and bounded for $\zeta \in \bar{D}_1 \cup \bar{D}_3$ and analytic in $D_1 \cup D_3$.
\item $A(\zeta)\overline{A(\bar{\zeta})} -  B(\zeta)\overline{B(\bar{\zeta})} = 1, \qquad \zeta \in \bar{D}_1 \cup \bar{D}_3$.
\item  $A(\zeta) = 1 + O\left(\frac{1}{\zeta}\right),
%+ O\left(\frac{e^{4i\zeta^4T}}{\zeta}\right), 
\qquad B(\zeta) = O\left(\frac{1}{\zeta}\right),
%+ O\left(\frac{e^{4i\zeta^4T}}{\zeta}\right), 
\qquad \zeta \to \infty, \quad \zeta \in D_1 \cup D_3.$
\end{itemize}

All of these properties follow from the analyticity and boundedness properties of $\mu_3(x,0,\zeta)$ and $\mu_1(0,t,\zeta)$, from the conditions of unit determinant, and from the large $\zeta$ asymptotics of these eigenfunctions.

\subsection{The Riemann-Hilbert problem}
Equations (\ref{seq}) and (\ref{Seq}) can be rewritten in a form expressing the jump condition of a $2 \times 2$ RH problem. This involves only tedious but straightforward algebraic manipulations. The final form is
\begin{equation}\label{Mjump}  
  M_-(x,t,\zeta) = M_+(x,t, \zeta)J(x,t,\zeta), \qquad \zeta \in \bar{D}_i \cap \bar{D}_j, \quad i,j = 1, \dots, 4,
\end{equation}
where the matrices $M_-$, $M_+$, and $J$ are defined as follows:
\begin{align}\label{MplusMminusdef}
M_+ = \left(\frac{\mu_2^{(1)}}{a(\zeta)}, \mu_3^{(12)}\right), \quad \zeta \in \bar{D}_1; \qquad
M_- = \left(\frac{\mu_1^{(2)}}{d(\zeta)}, \mu_3^{(12)}\right), \quad \zeta \in \bar{D}_2;
		\\ \nonumber
M_+ = \left(\mu_3^{(34)}, \frac{\mu_1^{(3)}}{\overline{d(\bar{\zeta})}}\right), \quad \zeta \in \bar{D}_3; \qquad
M_- = \left(\mu_3^{(34)}, \frac{\mu_2^{(4)}}{\overline{a(\bar{\zeta})}}\right), \quad \zeta \in \bar{D}_4;
\end{align}
\begin{equation}\label{ddef}
  d(\zeta) =a(\zeta)\overline{A(\bar{\zeta})} -  b(\zeta)\overline{B(\bar{\zeta})};
\end{equation}
\begin{equation}\label{Jdef}
J(x,t,\zeta) = \left\{ \begin{array}{ll}
J_1 & \zeta \in \bar{D}_1 \cap \bar{D}_2  \\
J_2 = J_3 J_4^{-1} J_1 & \zeta \in \bar{D}_2 \cap \bar{D}_3 \\
J_3 & \zeta \in \bar{D}_3 \cap \bar{D}_4 \\
J_4 & \zeta \in \bar{D}_4 \cap \bar{D}_1 \\
\end{array} \right.
\end{equation}
\begin{figure}
\begin{center}
   \includegraphics[width=.4\textwidth]{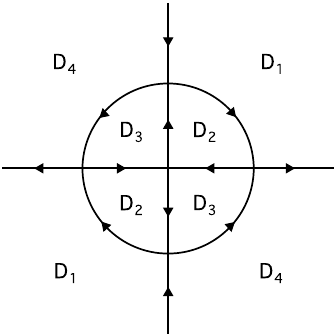} \\
     \begin{figuretext}\label{GNLSRH2.pdf}
        The contour for the Riemann-Hilbert problem in the complex $\zeta$-plane.
     \end{figuretext}
     \end{center}
\end{figure}
with
\begin{align}\label{J123def}
& J_1 = \begin{pmatrix} 1	&	0 	\\
\Gamma(\zeta)e^{2i\theta}	&	1 \end{pmatrix}, \qquad 
J_4 = \begin{pmatrix} 1	&	-\frac{b(\zeta)}{\overline{a(\bar{\zeta})}}e^{-2i\theta} 	\\
 \frac{\overline{b(\bar{\zeta})}}{a(\zeta)} e^{2i\theta}	&	\frac{1}{a(\zeta)\overline{a(\bar{\zeta})}} \end{pmatrix}, 
	\\ \nonumber
& J_3 = \begin{pmatrix} 1	&	- \overline{\Gamma(\bar{\zeta})}e^{-2i\theta} 	\\
0	&	1 \end{pmatrix};
\end{align}
\begin{equation}\label{Gammadef}
\theta(x,t,\zeta) = \zeta^2 x + \eta^2 t; \qquad \Gamma(\zeta) = \frac{\overline{B(\bar{\zeta})}}{a(\zeta)d(\zeta)}, \quad \zeta \in \bar{D}_2.
\end{equation}

The contour for this RH problem is depicted in Figure \ref{GNLSRH2.pdf}.

The matrix $M(x,t,\zeta)$ defined in (\ref{MplusMminusdef}) is in general a sectionally meromorphic function of $\zeta$. The possible poles of $M$ are generated by the zeros of $a(\zeta)$, of $d(\zeta)$, and by the complex conjugates of these zeros.
Since $a(\zeta)$ is an even function, each zero $\zeta_j$ of $a(\zeta)$ is accompanied by another zero at $\zeta_{j+1} = -\zeta_j$. Similarly, each zero $\lambda_j$ of $d(\zeta)$ is accompanied by a zero at $-\lambda_j$. Thus if $a(\zeta)$ has zeros, then necessarily it has an even number of zeros; similarly for $d(\zeta)$.

\begin{assumption}\label{zerosassumption}\upshape We assume that:
\item[(i)] The possible zeros of $a(\zeta)$ in $D_1$ and in $D_2$ are simple; these zeros are denoted by $\zeta_j$, where for $\{\zeta_j\}_{j = 1}^{2n_1}$, $\zeta_j \in D_1$, and for $\{\zeta_j\}_{j = 2n_1 + 1}^{2N}$, $\zeta_j \in D_2$.

\item[(ii)] The possible zeros of $d(\zeta)$ in $D_2$ are simple; these zeros are denoted by $\{\lambda_j\}_1^{2\Lambda}$. 

\item[(iii)] None of the zeros of $a(\zeta)$ coincides with a zero of $d(\zeta)$.

\item[(iv)]  $a(\zeta)$ and $d(\zeta)$ have no zeros on the contour of the RH problem.
\end{assumption}
In order to evaluate the associated residues we introduce the following notations:
\begin{itemize}
\item $[A]_1$ ($[A]_2$) denotes the first (second) column of a $2 \times 2$ matrix $A$. 

\item $\dot{a}(\zeta) = \frac{da}{d\zeta}$.

\item $M_j$ denotes the restriction of $M$ to $D_j$, $j = 1, \dots, 4$.

\item $\theta(\zeta_j) = \zeta_j^2 x + \eta_j^2 t$ and $\eta_j = \eta(\zeta_j)$.

\item $\text{Res}_{\bar{\lambda}_j} \, \overline{\Gamma(\bar{\zeta})}$ denotes the residue of the function $\zeta \mapsto \overline{\Gamma(\bar{\zeta})}$ at the pole $\zeta = \bar{\lambda}_j$.
\end{itemize}
We will now derive the following residue conditions:
\begin{align}\label{residue1}
\underset{\zeta_j}{\text{Res}} [M_1(x,t,\zeta)]_1 =& \frac{1}{\dot{a}(\zeta_j)b(\zeta_j)} e^{2i\theta(\zeta_j)} [M_1(x,t,\zeta_j)]_2, \qquad j = 1, \dots, 2n_1,
		\\\label{residue2}
\underset{\bar{\zeta}_j}{\text{Res}} [M_4(x,t,\zeta)]_2 =& \frac{1}{\overline{\dot{a}(\zeta_j)b(\zeta_j)}} e^{-2i\theta(\bar{\zeta}_j)} [M_4(x,t,\bar{\zeta}_j)]_1, \qquad j = 1, \dots, 2n_1,
	\\ \label{residue3}
\underset{\lambda_j}{\text{Res}} [M_2(x,t,\zeta)]_1 = &
\underset{\lambda_j}{\text{Res}} \, \Gamma(\zeta) \, e^{2i\theta(\lambda_j)} [M_2(x,t,\lambda_j)]_2, \qquad j = 1, \dots, 2\Lambda,
	\\\label{residue4}
\underset{\bar{\lambda}_j}{\text{Res}} [M_3(x,t,\zeta)]_2 =& 
\underset{\bar{\lambda}_j}{\text{Res}} \, \overline{\Gamma(\bar{\zeta})} \,
e^{-2i\theta(\bar{\lambda}_j)} [M_3(x,t,\bar{\lambda}_j)]_1, \qquad j = 1, \dots, 2\Lambda,
\end{align}
where (recall that $a(\zeta)$, by assumption \ref{zerosassumption}, satisfies $a(\lambda_j) \neq 0$)
$$\underset{\lambda_j}{\text{Res}}\,  \Gamma(\zeta) = \frac{\overline{B(\bar{\lambda}_j)}}{a(\lambda_j)\dot{d}(\lambda_j)},
\qquad
\underset{\bar{\lambda}_j}{\text{Res}}\, \overline{\Gamma(\bar{\zeta})}
= \frac{B(\bar{\lambda}_j)}{\overline{a(\lambda_j)\dot{d}(\lambda_j)} }.$$

In order to derive equation (\ref{residue1}) we note that the second column of equation (\ref{seq}) is
$$\mu_3^{(12)} = a\mu_2^{(4)} + b\mu_2^{(1)}e^{-2i\theta}.$$
Evaluating this equation at $\zeta = \zeta_j$, $j = 1, \dots, 2n_1$, we find
$$\mu_3^{(12)}(\zeta_j) = b(\zeta_j)\mu_2^{(1)}(\zeta_j)e^{-2i\theta(\zeta_j)},$$
where, for simplicity of notation, we have suppressed the $x$ and $t$ dependence. Thus, since $[M_1]_1 = \mu_2^{(1)}/a$, we find
$$\underset{\zeta_j}{\text{Res}} [M_1]_1 = \frac{\mu_2^{(1)}(\zeta_j)}{\dot{a}(\zeta_j)} =  \frac{e^{2i\theta(\zeta_j)} \mu_3^{(12)}(\zeta_j)}{\dot{a}(\zeta_j)b(\zeta_j)},$$
which is equation (\ref{residue1}).
The proof of (\ref{residue2}) is analogous.

In order to derive equation (\ref{residue3}) we note that the first column of the equation $M_2 = M_1J_1$ yields
\begin{equation}\label{M21M11}  
  [M_2]_1 = [M_1]_1 + \Gamma  e^{2i\theta} [M_1]_2.
\end{equation}
Since $[M_1]_2 = \mu_3^{(12)} = [M_2]_2$ and $[M_1]_1 = \mu_2^{(1)}/a$, it follows that each term in (\ref{M21M11}) has an analytic continuation for $\zeta \in D_2$; evaluating the residue at $\zeta = \lambda_j$, we find (\ref{residue3}).
Similarly, the second column of the equation $M_4 = M_3 J_3$ yields
\begin{equation}\label{M42M31}  
  [M_4]_2 =  -\overline{\Gamma(\bar{\zeta})} e^{-2i\theta} [M_3]_1 +  [M_3]_2.
\end{equation}
Since $[M_4]_2 = \mu_2^{(4)}/\overline{a(\bar{\zeta})}$, it follows that each term in (\ref{M42M31}) has an analytic continuation for $\zeta \in D_3$; evaluating the residue at $\zeta = \bar{\lambda}_j$, we find (\ref{residue3}).

\subsection{Reconstructing $u$}\label{inverseproblemsubsec}
The potential $u(x,t)$ can be reconstructed from the eigenfunctions $\mu_j(x,t,\zeta)$, $j = 1,2,3$, as follows. 
The second column of $\mu_2 = (\mu_2^{(1)}, \mu_2^{(4)})$ admits the expansion
$$\mu_2^{(4)} = \begin{pmatrix} 0 \\ 1 \end{pmatrix} + \frac{1}{\zeta}\begin{pmatrix} m(x,t) \\ n(x,t) \end{pmatrix} + O\left(\frac{1}{\zeta^2}\right), \quad \zeta \to \infty, \quad \zeta \in D_4,$$
where $m(x,t)$ and $n(x,t)$ are two functions independent of $\zeta$. Substituting this into the $x$-part of (\ref{mulax}) and considering terms of $O(\zeta)$ as $\zeta \to \infty$ in $D_4$, we infer that
\begin{equation}\label{recoverq}  
  u_x(x,t) = 2im(x,t)e^{2i\int^{(x,t)}_{(0, 0)} \Delta}.
\end{equation}
From equation (\ref{recoverq}) and its complex conjugate together with the conservation law
\begin{equation}\label{conservationlaw}   
  \left(u_xv_x\right)_t - \left(u_xv_x - uv\right)_x = 0,
\end{equation}
we obtain
$$u_xv_x = 4 |m|^2,\qquad u_xv_x - uv = - 4 \int_x^\infty  \left(|m|^2\right)_t dx'.$$
Thus, we are able to express the one-form $\Delta$ defined in (\ref{Deltadef}) in terms of $m$ as
\begin{align}\label{Deltarecover}
  \Delta = 2 |m|^2 dx -2 \left(\int_x^\infty  \left(|m|^2\right)_t dx'\right)dt.
\end{align}
The function $u$ can now be reconstructed as follows.
\begin{enumerate}
\item Compute $m$ according to
$$m(x,t) = \lim_{\zeta \to \infty} (\zeta \mu_2(x,t, \zeta))_{12}, \qquad \zeta \in D_4.$$
\item Determine $\Delta(x,t)$ from (\ref{Deltarecover}).
\item Compute $u(x,t)$ via
$$u(x,t) = -\int_x^\infty 2im(x',t)e^{2i\int^{(x',t)}_{(0, 0)} \Delta} dx'.$$
\end{enumerate}

\subsection{The global relation}
The spectral functions are not independent but satisfy an important global relation. Indeed, integrating the closed one-form $W = e^{i(\zeta^2 x + \eta^2 t)\hat{\sigma}_3} V\mu$ in (\ref{laxdiffform}) with $\mu = \mu_3$ around the boundary of the domain $\{0 < x < \infty, 0 < t < T_0\}$, we find
\begin{align}\label{squareintegration}
&\int_{\infty}^0 e^{i\zeta^2 x' \hat{\sigma}_3} (V_1\mu_3)(x', 0, \zeta)dx' 
+  \int_0^{T_0}  e^{i\eta^2 t' \hat{\sigma}_3}  (V_2\mu_3)(0, t', \zeta)dt'
		\\ \nonumber
&+ e^{i\eta^2 T_0 \hat{\sigma}_3} \int_0^\infty   e^{i\zeta^2 x' \hat{\sigma}_3}  (V_1\mu_3)(x', T_0, \zeta)dx'
= \lim_{X \to \infty} e^{i\zeta^2 X \hat{\sigma}_3} \int_0^{T_0}  e^{i\eta^2 t' \hat{\sigma}_3}  (V_2\mu_3)(X, t', \zeta)dt'.
\end{align}
Using that $s(\zeta) = \mu_3(0,0, \zeta)$ it follows from (\ref{mu3x0equation}) that the first term of this equation equals $s(\zeta) - I$. Equation (\ref{seq}) evaluated at $x = 0$ gives
$$\mu_3(0,t',\zeta) = \mu_2(0,t',\zeta)e^{-i\eta^2 t' \hat{\sigma}_3} s(\zeta).$$
Thus,
$$e^{i\eta^2 t' \hat{\sigma}_3}  (V_2\mu_3)(0, t', \zeta) = \left[e^{i\eta^2 t' \hat{\sigma}_3}  (V_2\mu_2)(0, t', \zeta)\right]s(\zeta).$$
This equation, together with (\ref{mu20tequation}), implies that the second term of (\ref{squareintegration}) is
$$\int_0^{T_0} e^{i\eta^2 t' \hat{\sigma}_3}  (V_2\mu_3)(0, t', \zeta)dt'  = [e^{i\eta^2 T_0 \hat{\sigma}_3} \mu_2(0,T_0, \zeta) - I]s(\zeta).$$
Hence, assuming that $u$ has sufficient decay as $x \to \infty$, equation (\ref{squareintegration}) becomes
\begin{equation}\label{squareintegration2} 
 -I + S(T_0, \zeta)^{-1}s(\zeta) + e^{i\eta^2 T_0 \hat{\sigma}_3} \int_0^\infty  e^{i\zeta^2 x' \hat{\sigma}_3}  (V_1\mu_3)(x', T_0, \zeta)dx'  = 0,
\end{equation}
where the first and second columns of this equation are valid for $\zeta^2$ in the lower and the upper half-plane, respectively, and $S(T_0, \zeta)$ is defined by
$$S(T_0, \zeta) = \left(e^{i\eta^2T_0\hat{\sigma}_3}\mu_2(0,T_0,\zeta)\right)^{-1}.$$
Letting $T_0 = T$ and noting that $S(\zeta) = S(T, \zeta)$, equation (\ref{squareintegration2}) becomes
$$ -I + S(\zeta)^{-1}s(\zeta) + e^{i\eta^2 T \hat{\sigma}_3} \int_0^\infty  e^{i\zeta^2 x' \hat{\sigma}_3}  (V_1\mu_3)(x', T, \zeta)dx'  = 0.$$
The $(12)$ component of this equation is the global relation
\begin{equation}\label{globalrelation}
   B(\zeta)a(\zeta) - A(\zeta)b(\zeta) = e^{2i\eta^2 T} c^+(\zeta), \qquad \zeta \in D_1,
\end{equation}
where
$$c^+(\zeta) = \int_0^\infty e^{2i\zeta^2 x' }  (V_1\mu_3)_{12}(x', T, \zeta)dx' .$$

\section{The spectral functions}\nequation
The above analysis motivates the following definitions for the spectral functions.

\begin{definition}[The spectral functions $a(\zeta)$ and $b(\zeta)$]\label{abdef}\upshape

Given $u_0(x) \in S(\R^+)$ in the Schwartz class, we define the map
$$\mathbb{S}:\{u_0(x)\} \to \{a(\zeta), b(\zeta)\}$$
by
$$\begin{pmatrix} b(\zeta) \\
 a(\zeta) \end{pmatrix} = [\mu_3(0,\zeta)]_2, \qquad \zeta \in \bar{D}_1 \cup \bar{D}_2,$$
where $\mu_3(x,\zeta)$ is the unique solution of the Volterra linear integral equation 
$$\mu_3(x,\zeta) = I + \int_{\infty}^{x} e^{i\zeta^2 (x' -x) \hat{\sigma}_3}V_1(x',0,\zeta) \mu_3(x',\zeta) dx',$$
and $V_1(x,0,\zeta)$ is given in terms of $u_0(x)$ by equation (\ref{V1initial}).
\end{definition}

\begin{proposition}\label{abprop}
The spectral functions $a(\zeta)$ and $b(\zeta)$ have the following properties:
\begin{enumerate}
\item[(i)] $a(\zeta)$ and $b(\zeta)$ are continuous and bounded for $\zeta \in \bar{D}_1 \cup \bar{D}_2$ and analytic in $D_1 \cup D_2$.
\item[(ii)] $a(\zeta) = 1 + O(1/\zeta), \quad b(\zeta) = O(1/\zeta), \qquad \zeta \to \infty, \quad \zeta \in D_1$.
\item[(iii)] $a(\zeta) = e^{-i\int_{(0,0)}^{(\infty, 0)}\Delta} + O(\zeta), \quad b(\zeta) = O(\zeta), \qquad \zeta \to 0, \quad \zeta \in D_2$.
\item[(iv)] $a(\zeta)\overline{a(\bar{\zeta})} -  b(\zeta)\overline{b(\bar{\zeta})} = 1, \qquad \zeta \in \bar{D}_1 \cup \bar{D}_2$.
\item[(v)] $a(-\zeta) = a(\zeta), \quad b(-\zeta) = -b(\zeta), \qquad \zeta \in \bar{D}_1 \cup \bar{D}_2$.
\item[(vi)] The map $\Q:\{a(\zeta), b(\zeta)\} \mapsto \{u_0(x)\}$, inverse to $\mathbb{S}$, is defined by
\begin{align}\label{recoverq0}
  u_0(x) =& -2i\int_x^\infty m(x')e^{4i\int^{x'}_0 |m(\xi)|^2 d\xi}dx', 
  	\\ \nonumber
  m(x) = &\lim_{\zeta \to \infty} \left(\zeta M^{(x)}(x, \zeta)\right)_{12}, \qquad \hbox{\upshape arg} \, \zeta \in [0, \pi/2],
\end{align}
where $M^{(x)}(x,\zeta)$ is the unique solution of the following RH problem
\begin{itemize}
\item $M^{(x)}(x,\zeta) = \left\{ \begin{array}{ll}
M_-^{(x)}(x,\zeta) &  \zeta \in D_3 \cup D_4 \\
M_+^{(x)}(x,\zeta) &  \zeta \in D_1 \cup D_2 \\
\end{array} \right.$ 

is a sectionally meromorphic function.

\item $M_-^{(x)}(x,\zeta) = M_+^{(x)}(x,\zeta) J^{(x)}(x,\zeta), \qquad \zeta^2 \in \R,$

where
\begin{equation}\label{Jxdef}
J^{(x)}(x, \zeta) = \begin{pmatrix} 1 & -\frac{b(\zeta)}{\overline{a(\bar{\zeta})}}e^{-2i\zeta^2x} \\
\frac{ \overline{b(\bar{\zeta})}}{a(\zeta)}e^{2i\zeta^2x}	& 	\frac{1}{a(\zeta)\overline{a(\bar{\zeta})}}\end{pmatrix}, \qquad \zeta^2 \in \R.
\end{equation}
\item $M^{(x)}(x,\zeta) = I + O\left(\frac{1}{\zeta}\right), \qquad \zeta \to \infty.$
\item $a(\zeta)$ may have $2N$ simple zeros $\{\zeta_j\}_{j = 1}^{2N}$ such that $\zeta_j \in D_1$, $j = 1, \dots, 2n_1$, and $\zeta_j \in D_2$, $j = 2n_1+1, \dots, 2N$. 
\item The possible simple poles of the first column of $M^{(x)}_+$ occur at $\zeta = \zeta_j$, $j = 1, \dots, 2N$, and the possible simple poles of the second column of $M^{(x)}_-$ occur at $\zeta = \bar{\zeta}_j$, $j = 1, \dots, 2N$.
The associated residues are given by
\begin{align}\label{xpartresidues}
\underset{\zeta_j}{\text{\upshape Res}} [M^{(x)}(x,\zeta)]_1 = \frac{e^{2i\zeta_j^2 x}}{\dot{a}(\zeta_j)b(\zeta_j)} [M^{(x)}(x, \zeta_j)]_2,  \qquad j = 1, \dots, 2N,
		\\ \label{xpartresiduesbar}
\underset{\bar{\zeta}_j}{\text{\upshape Res}} [M^{(x)}(x,\zeta)]_2 = \frac{ e^{-2i\bar{\zeta}_j^2 x}}{\overline{\dot{a}(\bar{\zeta}_j)b(\bar{\zeta}_j)}} [M^{(x)}(x, \bar{\zeta}_j)]_1,  \qquad j = 1, \dots, 2N.
\end{align}

\end{itemize}

\item[(vii)] We have
$$\mathbb{S}^{-1} = \Q.$$

\end{enumerate}

\end{proposition}
\proofbegin
$(i)-(ii)$ and $(iv)-(v)$ follow from the discussion in section \ref{abABsubsec}. $(iii)$ is proved in the appendix. We refer to the appendix of \cite{LDNLS} for a derivation of $(vi)$ and $(vii)$ in the similar case of the derivative NLS equation.
\proofend

\begin{definition}[The spectral functions $A(\zeta)$ and $B(\zeta)$]\label{ABdef}\upshape
Let $g_0(t)$ and $g_1(t)$ be smooth functions. The map
$$\tilde{\mathbb{S}}:\{g_0(t), g_1(t)\} \to \{A(\zeta), B(\zeta)\}$$
is defined by
$$\begin{pmatrix} B(\zeta) \\ A(\zeta) \end{pmatrix}
= [\mu_1(0, \zeta)]_2,$$
where $\mu_1(t, \zeta)$ is the unique solution of Volterra linear integral equation
$$\mu_1(t,\zeta) = I + \int_{T}^{t} e^{i\eta^2(t' - t)\hat{\sigma}_3} V_2(0,t',\zeta) \mu_1(t',\zeta) dt',$$
and $V_2(0,t,\zeta)$ is given in terms of $\{g_0(t), g_1(t)\}$ by equation (\ref{V2boundary}).
\end{definition}

\begin{proposition}\label{ABprop}
The spectral functions $A(\zeta)$ and $B(\zeta)$ have the following properties:
\begin{enumerate}
\item[(i)] $A(\zeta)$ and $B(\zeta)$ are continuous and bounded for $\zeta \in \bar{D}_1 \cup \bar{D}_3$ and analytic in $D_1 \cup D_3$. If $T < \infty$, then $A(\zeta)$ and $B(\zeta)$ are defined and analytic for all $\zeta \notin \{0, \infty\}$.
\item[(ii)] $A(\zeta) = 1 + O(1/\zeta), \quad B(\zeta) = O(1/\zeta), \qquad \zeta \to \infty, \quad \zeta \in D_1$.
\item[(iii)] $A(\zeta) = e^{-i\int_{(0,0)}^{(0, T)}\Delta} + O(\zeta), \quad B(\zeta) = O(\zeta), \qquad \zeta \to 0, \quad \zeta \in D_3$.
\item[(iv)] $A(\zeta)\overline{A(\bar{\zeta})} -  B(\zeta)\overline{B(\bar{\zeta})} = 1$, $\zeta \notin \{0, \infty\}$ ($\zeta \in \R \cup i\R \cup \{|\zeta| = 1/\sqrt{2}\}$ if $T = \infty$).
\item[(v)] $A(-\zeta) = A(\zeta), \quad B(-\zeta) = -B(\zeta)$.
\end{enumerate}

\end{proposition}
\proofbegin
$(iii)$ is proved in the appendix; the other properties follow from the discussion in section \ref{abABsubsec}.
\proofend

\begin{definition}[An admissible set of functions]\label{admissibledef}\upshape
Given $q_0(x) \in S(\R^+)$, define $a(\zeta)$ and $b(\zeta)$ according to Definition \ref{abdef}. Suppose that there exist smooth functions $g_0(t)$ and $g_1(t)$ such that

\begin{itemize}
\item The associated $A(\zeta)$ and $B(\zeta)$ defined according to Definition \ref{ABdef} satisfy the global relation
\begin{equation}\label{globaladmissible}
B(\zeta)a(\zeta) - A(\zeta)b(\zeta) = e^{2i\eta^2 T} c^+(\zeta), \qquad \zeta \in D_1 \cup D_2,
\end{equation}
where $c^+(\zeta)$ is analytic in $D_1 \cup D_2$, continuous and bounded for $\zeta \in \bar{D}_1$, and $c^+(\zeta) = O(1/\zeta)$ as $\zeta \to \infty$ in $D_1$.

\item The functions $u(x,0) = u_0(x)$, $u(0,t) = g_0(t)$, and $u_x(0,t) = g_1(t)$ are compatible with equation (\ref{GNLS2}) at $x = t = 0$, i.e. they satisfy
\begin{align*}
& g_0(0) = u_0(0), \qquad g_1(0) = u_0'(0),
	\\
& g_1'(0) + u_0(0) - 2iu_0'(0) - u_0''(0) +  i|u_0(0)|^2u_0'(0) = 0.
\end{align*}
\end{itemize}

Then we call $\{g_0(t), g_1(t)\}$ an {\it admissible set of functions with respect to $u_0(x)$.}
\end{definition}

\begin{remark}\upshape
If $T= \infty$ the functions $g_0(t)$ and $g_1(t)$ are assumed to belong to $S(\R^+)$ and the global relation (\ref{globaladmissible}) becomes
$$B(\zeta)a(\zeta) - A(\zeta)b(\zeta) = 0, \qquad \zeta \in \bar{D}_1.$$
\end{remark}

\section{The Riemann-Hilbert problem}\nequation

\begin{theorem}\label{RHtheorem}
Let $u_0(x) \in S(\R^+)$. Suppose that the functions $g_0(t)$ and $g_1(t)$ are admissible with respect to $u_0(x)$ (see Definition \ref{admissibledef}). Define the spectral functions $a(\zeta)$, $b(\zeta)$, $A(\zeta)$, and $B(\zeta)$ in terms of $u_0(x)$, $g_0(t)$, and $g_1(t)$ according to Definitions \ref{abdef} and \ref{ABdef}. Assume that the possible zeros $\{\zeta_j\}_{1}^{2N}$ of $a(\zeta)$ and $\{\lambda_j\}_{1}^{2\Lambda}$ of $d(\zeta)$ are as in Assumption \ref{zerosassumption}. Define $M(x,t,\zeta)$ as the solution of the following $2\times 2$ matrix RH problem:
\begin{itemize}
\item $M$ is sectionally meromorphic away from the boundaries of the $D_j$'s.

\item The possible poles of the first column of $M$ occur at $\zeta = \zeta_j$, $j = 1, \dots, 2n_1$, and $\zeta = \lambda_j$, $j = 1, \dots, 2\Lambda$. The possible poles of the second column of $M$ occur at $\zeta = \bar{\zeta}_j$,  $j = 1, \dots, 2n_1$, and $\zeta = \bar{\lambda}_j$, $j = 1, \dots, 2\Lambda$. The associated residues satisfy the relations in (\ref{residue1})-(\ref{residue4}).

\item $M$ satisfies the jump condition
$$M_-(x,t,\zeta) = M_+(x,t,\zeta)J(x,t,\zeta), \qquad \zeta \in \bar{D}_i \cap \bar{D}_j, \quad i,j = 1, \dots, 4,$$
where $M$ is $M_-$ for $\zeta \in D_2 \cup D_4$, $M$ is $M_+$ for $\zeta \in D_1 \cup D_3$, and $J$ is defined in terms of $a,b,A$, and $B$ by equations (\ref{ddef})-(\ref{Gammadef}), see Figure \ref{GNLSRH2.pdf}.

\item $M$ satisfies the normalization condition
\begin{equation}\label{Mnormalization}
  M(x,t,\zeta) = I + O\left(\frac{1}{\zeta}\right), \qquad \zeta \to \infty.
\end{equation}  
\end{itemize}
Then $M(x,t,\zeta)$ exists and is unique.

Define $u(x,t)$ in terms of $M(x,t,\zeta)$ by
\begin{equation}\label{recoverqxt}
  u(x,t) = -\int_x^\infty 2im(x',t)e^{2i\int^{(x',t)}_{(0, 0)} \Delta} dx', \qquad m(x,t) = \lim_{\zeta \to \infty} \left(\zeta M(x,t, \zeta)\right)_{12},
\end{equation}
$$\Delta = 2 |m|^2 dx -2  \left(\int_x^\infty  \left(|m|^2\right)_t dx'\right)dt.$$

Then $u(x,t)$ solves equation (\ref{GNLS2}). Furthermore,
\begin{equation}\label{initialboundaryvalues}  
  u(x,0) = u_0(x), \qquad u(0,t) = g_0(t), \qquad \text{and} \qquad u_x(0,t) = g_1(t).
\end{equation}
\end{theorem}
\proofbegin
In the case when $a(\zeta)$ and $d(\zeta)$ have no zeros, the unique solvability is a consequence of an appropriate vanishing lemma (cf. \cite{LDNLS} for a proof in the case of the derivative NLS).
If $a(\zeta)$ and $d(\zeta)$ have zeros this singular RH problem can be mapped to a regular one coupled with a system of algebraic equations \cite{F-I1996}.
Moreover, it follows from standard arguments using the dressing method \cite{Z-S1, Z-S2} that if $M$ solves the above RH problem and $u(x,t)$ is defined by (\ref{recoverqxt}), then $u(x,t)$ solves equation (\ref{GNLS2}). The proof that $u(0,t) = g_0(t)$ and $u_x(0,t) = g_1(t)$ follows arguments similar to the ones used in \cite{F-I-S}.
\proofend

\section{Linearizable boundary conditions}\label{linearizablesec}\nequation
It was shown in Theorem \ref{RHtheorem} that the solution $u(x,t)$ of equation (\ref{GNLS2}) can be expressed through the solution of a $2 \times 2$ matrix RH problem, which is uniquely characterized in terms of the spectral functions $a(\zeta)$, $b(\zeta)$, $A(\zeta)$, and $B(\zeta)$. The functions $a(\zeta)$ and $b(\zeta)$ are defined in terms of the initial data $u_0(x)$ through the solution of a linear Volterra integral equation (see Definition \ref{abdef}). However, the spectral functions $A(\zeta)$ and $B(\zeta)$ are, in general, not as readily obtained: The construction of $A(\zeta)$ and $B(\zeta)$ requires knowledge of both $g_0(t)$ and $g_1(t)$, whereas a boundary condition imposes only one condition on these functions; the second condition needed to determine $g_0(t)$ and $g_1(t)$ is the requirement that they satisfy the global relation (\ref{globalrelation}). For example, given the initial condition $u(x,0) = u_0(x)$ and the boundary condition $u(0, t) = g_0(t)$, in order to determine $A(\zeta)$ and $B(\zeta)$ according to Definition \ref{ABdef} we need to find a function $g_1(t)$ such that $\{g_0, g_1\}$ is an admissible set of functions with respect to $u_0$ (see Definition \ref{admissibledef}). In general, this problem involves solving a nonlinear Volterra integral equation cf. \cite{B-F-S, F2005}. 

However, for a particular class of boundary value problems it is possible to compute functions $\tilde{A}(\zeta)$ and $\tilde{B}(\zeta)$, which effectively replace $A(\zeta)$ and $B(\zeta)$ using only the algebraic manipulation of the global relation. More precisely, the solution $\tilde{M}$ of the RH problem involving $\tilde{A}, \tilde{B}$ instead of $A, B$, can be directly related to the solution to the original RH problem (see Lemma \ref{MtildeMlemma} below); hence $u(x,t)$ can be recovered from the large $\zeta$ asymptotics of $\tilde{M}$. When $T = \infty$ the functions $\tilde{A}$ and $\tilde{B}$ coincide with $A$ and $B$, so that the RH problems for $M$ and $\tilde{M}$ are identical. 
The class of boundary value problems which yield to this approach are referred to as linearizable. Thus, for a linearizable boundary condition the problem on the half-line can be solved as effectively as the problem on the line. 

\begin{theorem}\label{linearizableTh}
Fix $\alpha \in \R$ and let $u(x,t)$ satisfy equation (\ref{GNLS2}) with $T \leq \infty$, the initial condition 
$$u(x,0) = u_0(x), \qquad 0 < x < \infty,$$
and the boundary condition
$$u_x(0, t) = u(0, t) e^{i\alpha}, \qquad  0 \leq t < T.$$
Let
\begin{equation}\label{ndef}
n(\zeta) = \frac{e^{i\alpha} + 2i\zeta^2}{i + 2e^{i\alpha}\zeta^2}.
\end{equation}
Assume that the initial and boundary conditions are compatible at $(x,t) = (0,0)$. Furthermore, assume that
\begin{itemize}
\item[(i)] The possible zeros of $a(\zeta)$ in $D_1$ and $D_2$ denoted by $\{\zeta_j\}_{j = 1}^{2N}$ are simple, where $\zeta_j \in D_1$, $j = 1, \dots, 2n_1$, and $\zeta_j \in D_2$, $j = 2n_1+1, \dots, 2N$. 

\item[(ii)] The possible zeros of the function
\begin{equation}\label{mathcalDdef}  
  \mathcal{D}(\zeta) = a(\zeta)\overline{a\left(\frac{1}{2\bar{\zeta}}\right)} -  b(\zeta) \overline{b\left(\frac{1}{2\bar{\zeta}}\right)} n(\zeta), \qquad \zeta \in \bar{D}_1 \cup \bar{D}_2,
\end{equation}
in $D_2$, denoted by $\{\lambda_j\}_1^{2\Lambda}$, are simple.

\item[(iii)] None of the zeros of $a(\zeta)$ coincides with a zero of $\mathcal{D}(\zeta)$. 
\end{itemize}
Then the solution $u(x,t)$ is given by equation (\ref{recoverqxt}) with $M$ replaced by $\tilde{M}$, where $\tilde{M}$ is the solution of the Riemann-Hilbert problem in Theorem \ref{RHtheorem} with jump matrices and residue conditions defined by replacing $\Gamma$ in (\ref{J123def}) and (\ref{residue3})-(\ref{residue4}) with
\begin{equation}\label{tildeGammadef}
\tilde{\Gamma}(\zeta) = \frac{ n(\zeta) \overline{b(\frac{1}{2\bar{\zeta}})} }{a(\zeta) \mathcal{D}(\zeta)}, \qquad \zeta \in D_2.
\end{equation}
In the case when $e^{i \alpha} \neq \pm i$ the contour of the RH problem is deformed so as to avoid the zeros and poles of $n(\zeta)$, see remark \ref{deformremark}.
\end{theorem}
\begin{remark}\label{deformremark}
\upshape
  If $e^{i \alpha} \neq \pm i$, the function $n(\zeta)$ defined by (\ref{ndef}) has zeros and poles at 
  $$\zeta = \pm \frac{e^{i(\pi + 2 \alpha)/4}}{\sqrt{2}}  \qquad \hbox{and}\qquad \zeta = \pm \frac{e^{-i(\pi + 2 \alpha)/4}}{\sqrt{2}},$$
respectively. These zeros and poles lie on the contour separating $D_2$ and $D_3$ from $D_1$ and $D_4$. In what follows, we describe how this situation can be accomodated.

If $T = \infty$ it turns out that the RH problem of Theorem \ref{linearizableTh} is identical to the original RH problem (see remark \ref{Tinftyremark} below). Since we know that the jump matrices for this RH problem are singularity-free, it follows that all possible singularities induced by the zeros and poles of $n(\zeta)$ must cancel. 

If $T  < \infty$, since the jump matrices $J_2$ and $J_3$ defined in (\ref{J123def}) are analytic functions of $\zeta$ away from the zeros of $a$ and $d$ and their conjugates, we are permitted to make small deformations of the contour separating $D_2 \cup D_3$ from $D_1 \cup D_4$ in the RH problem. In order to avoid singularities on the contour in the formulation of the RH problem in Theorem \ref{linearizableTh}, we consider the modified RH problem obtained by deforming the contour so as to avoid the zeros and poles of $n(\zeta)$. We choose the deformations in such a way that the zeros and poles of $n(\zeta)$ lie in $D_1 \cup D_4$ with respect to the new contour. In the formulation of the deformed RH problem, the jump matrices defined in terms of the function $\tilde{\Gamma}(\zeta)$ given by (\ref{tildeGammadef}) are analytic on the contour. In this way, we also avoid the introduction of additional residue conditions for $\tilde{\Gamma}(\zeta)$ and $\overline{\tilde{\Gamma}(\bar{\zeta})}$, because $n(\zeta)$ has no poles in $D_2 \cup D_3$.
\end{remark}
\bigskip
{\it Proof of Theorem \ref{linearizableTh}. }
Recall that $A(\zeta)$ and $B(\zeta)$ are defined in terms of $\mu_2(t,\zeta)$. If $m(t,\zeta) = \mu_2(t,\zeta)e^{-i\eta^2 t \sigma_3}$, we have
%$$m(t,\zeta) = \begin{pmatrix} \overline{m_2(t, \bar{\zeta})}	&	m_1(t,\zeta)	\\
%\sigma \overline{m_2(t, \bar{\zeta})}	&	m_2(t, \zeta) \end{pmatrix}, \qquad
%m_1 = (\mu_2)_{12}e^{i\eta^2 t}, \qquad
%m_2 = (\mu_2)_{22}e^{i\eta^2 t}.$$
$$m_t + i\eta^2 \sigma_3 m = V_2(t,\zeta)m, \qquad m(0, \zeta) = I.$$
Since $\eta^2 = \left(\zeta - \frac{1}{2\zeta} \right)^2$ is invariant under $\zeta \to \frac{1}{2\zeta}$, the function $m(t, \frac{1}{2\zeta})$ satisfies the closely related equation
$$m_t + i\eta^2 \sigma_3 m = V_2 \left(t, \frac{1}{2\zeta}\right) m, \qquad m(0, \zeta) = I.$$
Suppose there exists a $t$-independent, nonsingular matrix $N(\zeta)$ such that
\begin{equation}\label{Nintertwine} 
 \left(i\eta^2\sigma_3 - V_2 \left(t, \frac{1}{2\zeta}\right)\right)N(\zeta) = N(\zeta) \left(i\eta^2\sigma_3 - V_2(t, \zeta) \right).
\end{equation}
Then
$$m\left(t, \frac{1}{2\zeta}\right) = N(\zeta)m(t, \zeta)N(\zeta)^{-1}.$$
This equation evaluated at $t = T$ yields 
\begin{equation}\label{SNSN}
  S\left(\frac{1}{2\zeta}\right)e^{i \eta^2 T \sigma_3} = N(\zeta)S(\zeta)e^{i \eta^2 T \sigma_3} N(\zeta)^{-1},
\end{equation}
which defines a relation between the spectral functions $A$ and $B$ evaluated at $\zeta$ and at $\frac{1}{2\zeta}$.

We note that a necessary condition for the existence of $N(\zeta)$ is that the determinants of the following two matrices are equal:
\begin{align*}
& i\eta^2\sigma_3 - V_2(t, \zeta) = e^{-i\int^t_0 \Delta_2dt' \hat{\sigma}_3} 
\begin{pmatrix} i(\zeta - \frac{1}{2\zeta})^2 + \frac{i}{2}u_xv_x	&	-(\zeta u_x + \frac{i}{2\zeta}u)	\\
-(\zeta v_x - \frac{i}{2\zeta}v)					&	-i(\zeta - \frac{1}{2\zeta})^2 - \frac{i}{2}u_xv_x \end{pmatrix},
	\\
& i\eta^2\sigma_3 - V_2\left(t, \frac{1}{2\zeta}\right) = e^{-i\int^t_0 \Delta_2dt' \hat{\sigma}_3} 
\begin{pmatrix} i(\zeta - \frac{1}{2\zeta})^2 + \frac{i}{2}u_xv_x	&	-(\frac{1}{2\zeta} u_x + i \zeta u)	\\
-(\frac{1}{2\zeta} v_x - i \zeta v)					&	-i(\zeta - \frac{1}{2\zeta})^2 - \frac{i}{2}u_xv_x \end{pmatrix}.
\end{align*}
This implies that $|u_x|^2 = |u|^2$, i.e. $u_x(0, t) = u(0, t) e^{i\alpha}$ for some $\alpha(t) \in \R.$ Assume that $\alpha$ is a constant, then equation (\ref{Nintertwine}) is satisfied with 
$$N(\zeta) = \begin{pmatrix} i - 2e^{-i\alpha}\zeta^2 	&	0	\\
0	&	- e^{-i\alpha} + 2i\zeta^2 \end{pmatrix}.$$
Thus, with $n(\zeta)$ defined according to (\ref{ndef}), the second column of equation (\ref{SNSN}) yields
\begin{equation}\label{ABsymmetry}
  A\left(\frac{1}{2\zeta}\right) = A(\zeta), \qquad B\left(\frac{1}{2\zeta} \right) = n(\zeta) B(\zeta), \qquad \zeta \in D_1 \cup D_3.
\end{equation}
Also, observe that
\begin{equation}\label{nsymmetries}
  n\left(\frac{1}{2\zeta}\right) = \frac{1}{n(\zeta)} = \overline{n(\bar{\zeta})}.
\end{equation}

Letting $\zeta \to \frac{1}{2\zeta}$ in the definition of $\overline{d(\bar{\zeta})}$, i.e. in the equation
$$\overline{d(\bar{\zeta})} = \overline{a(\bar{\zeta})}A(\zeta) -  \overline{b(\bar{\zeta})}B(\zeta), \qquad \zeta \in D_3,$$
we find, in view of the symmetry (\ref{ABsymmetry}),
\begin{equation}\label{d12barzetaeq}
\overline{d\left(\frac{1}{2\bar{\zeta}}\right)} = \overline{a\left(\frac{1}{2\bar{\zeta}}\right)}A(\zeta) -  \overline{b\left(\frac{1}{2\bar{\zeta}}\right)}n(\zeta) B(\zeta), \qquad \zeta \in D_1.
\end{equation}
Equation (\ref{d12barzetaeq}) together with the global relation
$$B(\zeta)a(\zeta) - A(\zeta)b(\zeta) = e^{2i\eta^2 T} c^+(\zeta), \qquad \zeta \in D_1,$$
are two algebraic relations for $A$ and $B$. The solution of these equations is
\begin{equation}\label{ABinD1}
\begin{cases}
A(\zeta) = \frac{1}{\mathcal{D}(\zeta)}\left(a(\zeta)\overline{d\left(\frac{1}{2\bar{\zeta}}\right)} +  \overline{b\left(\frac{1}{2\bar{\zeta}}\right)}n(\zeta)e^{2i\eta^2 T} c^+(\zeta)\right)
	\\ 
B(\zeta) = \frac{1}{\mathcal{D}(\zeta)}\left(b(\zeta)\overline{d\left(\frac{1}{2\bar{\zeta}}\right)} + \overline{a\left(\frac{1}{2\bar{\zeta}}\right)}e^{2i\eta^2 T} c^+(\zeta)\right)
\end{cases} \qquad \zeta \in D_1,
\end{equation}
where $\mathcal{D}(\zeta)$ is given by (\ref{mathcalDdef}). Letting $\zeta \to \frac{1}{2 \zeta}$ in (\ref{ABinD1}) and using (\ref{ABsymmetry}), we find
\begin{equation}\label{ABinD3}
\begin{cases}
A(\zeta) = \frac{1}{\mathcal{D}\left(\frac{1}{2\zeta}\right)}\left(a\left(\frac{1}{2\zeta}\right)\overline{d\left(\bar{\zeta}\right)} +  \overline{b\left(\bar{\zeta}\right)}n\left(\frac{1}{2\zeta}\right) e^{2i\eta^2 T} c^+ \left(\frac{1}{2\zeta}\right)\right)
	\\ 
B(\zeta) = \frac{1}{n(\zeta) \mathcal{D} \left(\frac{1}{2\zeta}\right)}\left(b\left(\frac{1}{2\zeta}\right)\overline{d\left(\bar{\zeta}\right)} + \overline{a\left(\bar{\zeta}\right)}e^{2i\eta^2 T} c^+\left(\frac{1}{2\zeta}\right)\right)
\end{cases} \qquad \zeta \in D_3.
\end{equation}

The right-hand sides of equations (\ref{ABinD1}) and (\ref{ABinD3}) involve the unknown functions $c^+$ and $d$. However, it turns out that it is possible to pose an equivalent RH problem for which $A$ and $B$ are replaced by 
\begin{equation}\label{tildeABdef}
\tilde{A}(\zeta) = \begin{cases}
\frac{a(\zeta)\overline{d\left(\frac{1}{2\bar{\zeta}}\right)}}{\mathcal{D}(\zeta)}, & \zeta \in D_1,
	\\
	\\
\frac{a\left(\frac{1}{2\zeta}\right)\overline{d\left(\bar{\zeta}\right)}}{\mathcal{D}\left(\frac{1}{2\zeta}\right)}, & \zeta \in D_3,
\end{cases}
\qquad 
\tilde{B}(\zeta) = 
\begin{cases}
\frac{b(\zeta)\overline{d\left(\frac{1}{2\bar{\zeta}}\right)}}{\mathcal{D}(\zeta)}, & \zeta \in D_1,
	\\
	\\
	\frac{b\left(\frac{1}{2\zeta}\right)\overline{d\left(\bar{\zeta}\right)} }{n(\zeta) \mathcal{D} \left(\frac{1}{2\zeta}\right)}, & \zeta \in D_3.
\end{cases}
\end{equation}
Before describing this new RH problem and its relation to the original RH problem, we show that the function 
$$\tilde{\Gamma}(\zeta) = \frac{\overline{\tilde{B}\left(\bar{\zeta}\right)}}{a(\zeta)\tilde{d}(\zeta)},$$
defined by replacing $A, B$ with $\tilde{A}, \tilde{B}$ in the definition (\ref{Gammadef}) of $\Gamma$, can be written as in (\ref{tildeGammadef}).
Indeed, note that
$$\frac{\tilde{B}(\zeta)}{\tilde{A}(\zeta)} = \frac{b\left(\frac{1}{2\zeta}\right)}{n(\zeta)a\left(\frac{1}{2\zeta}\right)},  \qquad \zeta \in D_3.$$
Therefore,
\begin{equation}\label{tildeBtildeAbna}
\frac{\overline{\tilde{B}(\bar{\zeta})}}{\overline{\tilde{A}(\bar{\zeta})}} = \frac{\overline{b\left(\frac{1}{2\bar{\zeta}}\right)}}{\overline{n(\bar{\zeta})}\overline{a\left(\frac{1}{2\bar{\zeta}}\right)}}, \qquad \zeta \in D_2.
\end{equation}
It follows, using (\ref{nsymmetries}), that
$$\tilde{\Gamma}(\zeta) = \frac{ \overline{\tilde{B}(\bar{\zeta})}/\overline{\tilde{A}(\bar{\zeta})}}{a(\zeta)  (a(\zeta) -  b(\zeta)\overline{\tilde{B}(\bar{\zeta})}/\overline{\tilde{A}(\bar{\zeta})})}
= \frac{ n(\zeta) \overline{b(\frac{1}{2\bar{\zeta}})} }{a(\zeta) \mathcal{D}(\zeta)}, \qquad \zeta \in D_2,$$
and the result follows.

We now consider how the replacements $A \to \tilde{A}$ and $B \to \tilde{B}$ affect the RH problem of Theorem \ref{RHtheorem}. The following lemma is proved in the same way as Proposition 3.1 in \cite{F2002}.

\begin{lemma}\label{MtildeMlemma}
Let $\tilde{J}_i$, $i = 1, \dots, 4$, be the jump matrices defined according to (\ref{J123def}) with $\Gamma$ replaced by $\tilde{\Gamma}$, which is given by (\ref{tildeGammadef}). Let $\tilde{M}(t,x,\zeta)$ satisfy a RH problem similar to that of $M(x,t,\zeta)$ but with jump matrices $J_i$, $i = 1, \dots, 4$, replaced by $\tilde{J}_i$, $i = 1, \dots, 4$, and the function $\Gamma$ in the residue conditions (\ref{residue3})-(\ref{residue4}) replaced by $\tilde{\Gamma}$.
Then the solutions $M$ and $\tilde{M}$ are related by
\begin{equation}\label{MtildeMrelations}
M_1 = \tilde{M}_1, \quad 
M_2 = \tilde{M}_2 \tilde{J}_1^{-1} J_1, \quad 
M_3 = \tilde{M}_3 \tilde{J}_3 J_3^{-1}, \quad 
M_4 = \tilde{M}_4.
\end{equation}
\end{lemma}

\bigskip
We can now finish the proof of Theorem \ref{linearizableTh}.
From Lemma \ref{MtildeMlemma} it follows that $\tilde{M}_1 = M_1$. Hence the solution $u(x,t)$ is given by equation (\ref{recoverqxt}) with $M$ replaced by $\tilde{M}$. This completes the proof of Theorem \ref{linearizableTh}.
\proofend

\begin{remark}\label{Tinftyremark}\upshape
In the case when $T = \infty$ the proof of Theorem \ref{linearizableTh} simplifies considerably, because in this case the RH problems for $M$ and $\tilde{M}$ are identical. Indeed, when $T = \infty$ the global relation (\ref{globalrelation}) is
\begin{equation}\label{globalagain}
  B(\zeta)a(\zeta) - A(\zeta)b(\zeta) = 0, \qquad \zeta \in D_1.
\end{equation}
Hence we can set $c^+(\zeta) = 0$ in equations (\ref{ABinD1}) and (\ref{ABinD3}), which implies that $A = \tilde{A}$ and $B = \tilde{B}$.
However, the case $T < \infty$ is important because it allows solutions $u(x,t)$ for which $u(0,t)$ does not decay to zero as $t \to \infty$. We will see explicit examples where this occurs in the following sections.
\end{remark}

\section{One-soliton restricted to the half-line}\label{solitonsec}\nequation
The rest of the paper is devoted to the analysis of particular examples of IBV problems which admit explicit solutions. These examples illustrate the methods described earlier and provide a direct check on the formalism. In particular, we will consider problems which satisfy linearizable boundary conditions, and for which the approach described in the previous section can be implemented. For these examples, we will be able to see in detail how the RH problem of Theorem \ref{linearizableTh} involving $\tilde{\Gamma}$ is related to the RH problem involving $\Gamma$. 

When viewed on the real line, equation (\ref{GNLS2}) admits a four-parameter family of one-soliton solutions \cite{L-F}: for any values of the parameters
$$\gamma \in (0, \pi), \qquad \Delta_0 > 0, \qquad \Sigma_0 \in \R, \qquad x_0 \in \R,$$
the function
\begin{equation}\label{onesoliton}
u(x,t) =  -\frac{2i\sin \gamma}{\Delta_0}\frac{e^{-  i \gamma}e^{-2i\Sigma(x,t)}}{e^{2\theta(x,t)} + e^{ i \gamma}e^{-2\theta(x,t)}}, \qquad x\in \R, \ t > 0,
\end{equation}
where 
$$\Sigma(x,t) = - t -\Delta_0^2 \cos \gamma\left( x +  \left(1 + \frac{1}{4\Delta_0^4}\right)t\right) + \Sigma_0$$
and
$$\theta(x,t) = \Delta_0^2 \sin \gamma \left(x  - x_0 +  \left(1 - \frac{1}{4\Delta_0^4}\right)t\right),$$ 
is a one-soliton solution of equation (\ref{GNLS2}). 
According to Theorem \ref{linearizableTh} $u(x,t)$ satisfies a linearizable boundary condition provided that there exists an $\alpha \in \R$ such that $u_{x}(0,t) = e^{i \alpha} u(0,t)$ for $0 < t < T$. For any values of the parameters $\gamma, \Sigma_0, x_0$, it follows from (\ref{onesoliton}) that
$$\left|\frac{u_{x}(x,t)}{u(x,t)}\right|^2 = 4 \Delta_0^4, \qquad x\in \R, \ t > 0.$$
Since by assumption $\Delta_0 > 0$, we infer that $u(x,t)$ satisfies a linearizable boundary condition provided that $\Delta_0 = \frac{1}{\sqrt{2}}$. Henceforth we restrict ourselves to this case; the one-soliton solutions satisfying this restriction will be denoted by $u^s(x,t)$.

\subsection{The one-form $\Delta$}
For $u^s(x,t)$ given by (\ref{onesoliton}) with $\Delta_0 = \frac{1}{\sqrt{2}}$, we can compute the one-form $\Delta$ defined by  
(\ref{Deltadef}). A calculation shows that
$$|u^s(x,t)|^2 = |u_x^s(x,t)|^2 = \left(4 \arctan\left(\tan \left(\frac{\gamma }{2}\right) \tanh
   ((x-x_0) \sin{\gamma})\right)\right)_x.$$
Thus
\begin{equation}\label{Deltasolitonint}
\Delta(x,t) = d\left(2 \arctan\left(\tan \left(\frac{\gamma }{2}\right) \tanh((x-x_0) \sin{\gamma})\right)\right).
\end{equation}
From (\ref{Deltasolitonint}) we derive explicit expressions for the quantity of interest, $e^{ i\int_{-\infty}^{x} \Delta}$.\footnote{Here and in the sequel we only specify the $x$ coordinates in the limits of integration when integrating $\Delta$, because the $dt$ component of $\Delta$ vanishes.}
In this respect, it is more convenient to first compute the expression
$$e^{ i \int_{-\infty}^\infty \Delta} = e^{2 i \gamma },$$
and then to use the identity (cf. \cite{L-F})
$$e^{ i\int_{-\infty}^{x} \Delta} = \frac{e^{4\theta(x,t)} + e^{-i \gamma}}{e^{4\theta (x,t)} + e^{i \gamma}} e^{i\int_{-\infty}^\infty \Delta}.$$
In particular, this yields
\begin{equation}\label{integralidentity}
e^{i \int_{-\infty}^0 \Delta} = \frac{1+e^{i \gamma -2 x_0 \sin{\gamma}}}{1+e^{-i \gamma -2 x_0 \sin{\gamma}}}.
\end{equation}

\subsection{Eigenfunctions on the line}
Applying the standard inverse scattering formalism, we may derive explicit formulas for the eigenfunctions on the line corresponding to $u^s(x,0)$. As in (\ref{sSandabAB}), we let $a^{line}(\zeta) = (s^{line}(\zeta))_{22}$ (since $u^s(x,t)$ is a soliton solution the spectral function $b^{line}(\zeta)= (s^{line}(\zeta))_{12}$ vanishes identically), where $s^{line}(\zeta)$ is given by (see \cite{L-F})
\begin{equation}\label{slineeq} 
  s^{line}(\zeta) = \lim_{x \to -\infty} e^{i\zeta^2 x\hat{\sigma}_3} \mu_2^{line}(x,\zeta).
\end{equation}
Here the eigenfunction $\mu_2^{line}(x,\zeta)$ is defined by the integral equation
$$\mu_2^{line}(x,\zeta) = I - \int_x^{\infty} e^{i \zeta^2 (x' - x) \hat{\sigma}_3} (V_1^{line} \mu_2^{line})(x',\zeta)dx',$$
where
\begin{equation}\label{V1linehalfrelation} 
  V_1^{line} = e^{- i \int_{-\infty}^0 \Delta \hat{\sigma}_3} V_1.
\end{equation}
The soliton $u^s(x,t)$ is characterized by the position of the zeros $\zeta_1^{line} = - \zeta_2^{line} \in D_1 \cup D_2$ of $a^{line}(\zeta)$ and the value of a normalization constant $C_1 \in \C$. These are related to the four parameters $\Delta_0 > 0, \gamma \in (0, \pi), x_0 \in \R,$ and $\Sigma_0 \in \R$, by
\begin{equation}\label{parameterrelation}
\left(\zeta_1^{line}\right)^2 = \Delta_0^2(-\cos \gamma + i\sin \gamma), \qquad C_1 = i \Delta_0 \sin \gamma e^{4 i \gamma } e^{2i\Sigma_0}e^{2\Delta_0^2 x_0 \sin\gamma }.
\end{equation}

It is described in \cite{L-F} how to find $u(x,t)$ from the solution $M^{line}(x, \zeta)$ of a $2 \times 2$ matrix RH problem. We will use the following facts derived in \cite{L-F} (in order to simplify the notation we will in the rest of this subsection write $z_j$ instead of $\zeta_j^{line}$, $j =1,2$):
\begin{itemize}
\item $M^{line}$ satisfies the algebraic system 
\begin{align} \label{Msolitonsystem} 
&\overline{M_{22}^{line}(x, z_1)} = 1 + C_1 e^{2iz_1^2 x}\left(\frac{1}{\bar{z}_1 - z_1} - \frac{1}{\bar{z}_1 + z_1}\right) M_{12}^{line}(x, z_1)
	\\
&\overline{M_{12}^{line}(x, z_1)} = C_1 e^{2iz_1^2 x} \left(\frac{1}{\bar{z}_1 - z_1}+ \frac{1}{\bar{z}_1 + z_1}\right)M_{22}^{line}(x,z_1).
\end{align}

\item $M^{line}$ satisfies the identity
\begin{equation}\label{Msolitonidentity}
\begin{pmatrix} \overline{M_{22}^{line}(x, \bar{\zeta})} \\ \overline{M_{12}^{line}(x, \bar{\zeta})} \end{pmatrix} = \begin{pmatrix} 1	\\	0 \end{pmatrix} + \sum_{j = 1}^{2} \frac{C_j e^{2iz_j^2 x}}{\zeta - z_j}\begin{pmatrix} M_{12}^{line}(x, z_j) \\ M_{22}^{line}(x, z_j) \end{pmatrix}.
\end{equation}

\item The eigenfunction $\mu_2^{line}$ can be obtained from $M^{line}$ via
\begin{equation}\label{Mlinemu2line}
  [M^{line}(x,\zeta)]_2 = [\mu_2^{line}(x,\zeta)]_2, \qquad \zeta \in D_1 \cup D_2.
\end{equation}
\end{itemize}
We deduce from (\ref{Msolitonsystem}) that
\begin{align*}
  &\overline{M_{22}^{line}(x,z_1)} =  \frac{e^{2 i x \bar{z}_1^2}
   \left(z_1^2-\bar{z}_1^2\right)^2}{4 C_1
   \bar{C}_1 e^{2 i x z_1^2} z_1^2+e^{2 i x
   \bar{z}_1^2} \left(z_1^2-\bar{z}_1^2\right)^2},
   	\\
   &\overline{M_{12}^{line}(x,z_1)} =  \frac{2 C_1 \bar{z}_1
   \left(\bar{z}_1^2-z_1^2\right)}{4 C_1
   \bar{C}_1 e^{-2 i x \bar{z}_1^2} \bar{z}_1^2+e^{-2 i x
   z_1^2} \left(\bar{z}_1^2-z_1^2\right)^2}.
\end{align*}  
Substituting this into (\ref{Msolitonidentity}), we find 
\begin{align*}
  & M_{12}^{line}(x,\zeta) = \frac{\bar{C}_1 \left(z_1^2-\bar{z}_1^2\right)^2
   \left(\frac{1}{\bar{z}_1+\zeta }+\frac{1}{\zeta
   -\bar{z}_1}\right)}{4 C_1 \bar{C}_1 e^{2 i x
   z_1^2} z_1^2+e^{2 i x \bar{z}_1^2}
   \left(z_1^2-\bar{z}_1^2\right)^2},
   	\\
   & M_{22}^{line}(x,\zeta) = \frac{4 C_1 \bar{C}_1 e^{2 i x z_1^2}
   \left(z_1^2-\bar{z}_1^2\right)
   \bar{z}_1^2}{\left(4 C_1 \bar{C}_1 e^{2 i x
   z_1^2} \bar{z}_1^2+e^{2 i x \bar{z}_1^2}
   \left(z_1^2-\bar{z}_1^2\right)^2\right)
   \left(\bar{z}_1^2-\zeta ^2\right)}+1.
\end{align*}
Now (\ref{Mlinemu2line}) yields an explicit expression for $[\mu_2^{line}(x,\zeta)]_2$. The first column of $\mu_2^{line}$ is obtained by symmetry. The complimentary eigenfunction $\mu_1^{line}$ can be found by analogous arguments, but will not be needed in what follows.

\subsection{Spectral functions on the half-line}
In view of the relation (\ref{V1linehalfrelation}) between $V_1^{line}$ and $V_1$, we deduce that
\begin{equation}\label{mu3frommu2line}
  [\mu_3(x,0,\zeta)]_2 = \begin{pmatrix} e^{2 i \int_{-\infty}^0 \Delta} (\mu_2^{line}(x,\zeta))_{12} \\ (\mu_2^{line}(x,\zeta))_{22} \end{pmatrix}.
\end{equation}
Indeed, the column vector on the right-hand side approaches $\begin{pmatrix} 0 \\ 1 \end{pmatrix}$ as $x \to \infty$ and satisfies the second column of the equation
$$\mu_{x} + i\zeta^2[\sigma_3, \mu] = V_1 \mu.$$
The spectral functions $a(\zeta)$ and $b(\zeta)$ are given by
$$\begin{pmatrix} b(\zeta) \\ a(\zeta) \end{pmatrix} = [\mu_3(0, 0, \zeta)]_2.$$
We find (as above $z_1$ denotes $\zeta_1^{line}$)
\begin{align}
& a(\zeta) = \frac{4 C_1 \bar{C}_1
   \left(z_1^2-\bar{z}_1^2\right)
   \bar{z}_1^2}{\left(4 C_1 \bar{C}_1
   \bar{z}_1^2+\left(z_1^2-\bar{z}_1^2\right)^2\right) \left(\bar{z}_1^2-\zeta^2\right)}+1,
   	\\
& b(\zeta) = \frac{\bar{C}_1 e^{2 i \int_{-\infty}^0 \Delta}
   \left(z_1^2-\bar{z}_1^2\right)^2
   \left(\frac{1}{\bar{z}_1+\zeta }+\frac{1}{\zeta -\bar{z}_1}\right)}{4 C_1 \bar{C}_1
   z_1^2+\left(z_1^2-\bar{z}_1^2\right)^2}.
\end{align}
Using (\ref{integralidentity}) and writing this in terms of $\gamma$, $x_0$, and $\Sigma_0$, we find 
\begin{align}\label{aexplicit}
& a(\zeta) =   \frac{2 \zeta ^2+e^{2 x_0 \sin{\gamma}}+\left(2 e^{2 x_0
   \sin{\gamma}} \zeta ^2+1\right) \cos{\gamma}+i \left(2 e^{2
   x_0 \sin{\gamma}} \zeta ^2+1\right) \sin{\gamma}}{\left(2
   \zeta ^2+\cos{\gamma}+i \sin{\gamma}\right) \left(e^{2 x_0
   \sin{\gamma}} \cos{\gamma}+i e^{2 x_0 \sin{\gamma}} \sin
  {\gamma}+1\right)},	\\
& b(\zeta) = \frac{2 \sqrt{2} e^{x_0 \sin{\gamma}-2 i (\gamma +\Sigma_0)} \left(e^{i \gamma }+e^{2 x_0 \sin{\gamma}}\right)^2
   \zeta  \sin{\gamma}}{\left(1+e^{i \gamma +2 x_0 \sin{\gamma}}\right)^2 \left(2 \zeta ^2+e^{i \gamma }\right) \left(e^{2 x_0
   \sin{\gamma}} (i \cos{\gamma}+\sin{\gamma})+i\right)}.
\end{align}
Note that $b(\zeta)$ is nonzero in contrast to the case on the full line.

\subsection{The zeros of $a(\zeta)$}
The function $a(\zeta)$ as given in (\ref{aexplicit}) has simple zeros at $\zeta_1$ and $\zeta_2 = - \zeta_1$, where
$$\zeta_1^2 = -\frac{1}{2} \cos \left(\frac{1}{2} (\gamma +2 i x_0 \sin{\gamma})\right) \sec \left(\frac{1}{2} (\gamma -2 i x_0 \sin {\gamma})\right).$$   
Note that
$$\lim_{x_0 \to \infty} \zeta_1^2 = \frac{1}{2} (i \sin{\gamma}-\cos{\gamma}) = (\zeta_1^{line})^2.$$
Thus, in this limit the zeros of $a(\zeta)$ approach those of $a^{line}(\zeta)$ as expected: the initial profile of $u(x,t)$ is localized around $x_0$, so that as $x_0 \to \infty$ the effect of the boundary becomes negligible. Another manifestation of the negligible influence of the boundary in this limit is that 
$$\lim_{x_0 \to \infty} b(\zeta) = 0.$$
Furthermore, a straightforward computation yields
$$|\zeta_1| = \frac{1}{\sqrt{2}}.$$ 
This equation implies that the zeros lie {\it on} the contour of the RH problem and therefore the explicit solutions studied in this section fall slightly outside the theoretical framework presented earlier (where it was assumed that the zeros do {\it not} lie on the contour). In the following section we will see how to modify the formalism developed earlier in order to accomodate this situation.

Let us also point out that
$$\zeta_1^2 = \frac{i}{2} e^{i \alpha}.$$

As a final remark we mention that $\zeta_1^2$ lies in the upper half plane if and only if $x_0 \geq 0$. Therefore, $a(\zeta)$ has its zeros in the first and third quadrants if and only if $x_0 \geq 0$.

\section{An explicit example: direct approach}\label{exampledirectsec}\nequation
In the previous section we considered the IBV problem satisfied by the restriction of the one-soliton solution $u^s(x,t)$ given by (\ref{onesoliton}) with $\Delta_0 = \frac{1}{\sqrt{2}}$ to the half-line. The formulas derived from $u^s(x,t)$ simplify considerably when, in addition to setting $\Delta_0 = \frac{1}{\sqrt{2}}$, we also make some special choices for the parameters $\gamma, x_0$, and $\Sigma_0$. In the following two sections for simplicity we let
\begin{equation}\label{parameterchoice}  
  \gamma = \frac{\pi}{2}, \qquad x_0 = 0, \qquad \Sigma_0 = 0.
\end{equation}
This yields the following simple solution denoted by $u^p(x,t)$:
\begin{equation}\label{simplestu}
  u^p(x,t) = -\frac{2 \sqrt{2} e^{2 i t+x}}{i+e^{2 x}}.
\end{equation}
Hence, we will consider the following IBV problem:
\begin{equation}\label{exampleIBVproblem}
\begin{cases}
  u(x,t) \hbox{  satisfies (\ref{GNLS2}) on  } \{x>0, t>0\},
  	\\
  u_0(x) = -\frac{2 \sqrt{2} e^x}{i+e^{2 x}}, \quad x > 0,	 \qquad  u_x(0,t) =  i u(0,t), \quad t > 0.
\end{cases}
\end{equation}
We first derive, using a direct approach, explicit expressions for the eigenfunctions $\mu_j$, $j = 1, 2,3$, as well as for the jump matrices $J_i$, $i = 1, \dots, 4$. We then construct the solution $M(x,t,\zeta)$ of the RH problem and verify that it indeed satisfies the appropriate jump and residue conditions. One nontrivial aspect of this example is that the zeros of $a(\zeta)$ and $d(\zeta)$ coincide and lie on the contour of the RH problem. It will however become clear how to deal with this subtlety. We will also see explicitly how the solution $u(x,t)$ can be recovered from the large $\zeta$ asymptotics of $M(x,t,\zeta)$. 
Subsequently, in the next section, we use the linearizable approach, as described in section \ref{linearizablesec}, to analyze the same problem.

\subsection{The one-form $\Delta$}
Since
$$|u^p(x,t)|^2 = |u_x^p(x,t)|^2 =\frac{8 e^{2 x}}{1+e^{4 x}},$$
we find that the one-form $\Delta$ is given by
 \begin{equation}\label{Deltaexample}  
  \Delta(x,t) = \frac{1}{2}|u_x^p|^2 dx = d\left(2 \arctan{e^{2 x}}\right).
\end{equation}
Therefore,
\begin{equation}\label{Deltaints}
\int_{-\infty}^{0} \Delta = \frac{\pi}{2}, \qquad \int_{-\infty}^{\infty} \Delta = \pi, \qquad \int_{0}^{x} \Delta = 2 \arctan(e^{2 x}) - \frac{\pi}{2}.
\end{equation}
Note that
$$e^{2 i \int_{0}^{x} \Delta} = \frac{2}{(\cosh (x)-i \sinh (x))^2}-1.$$

\subsection{Eigenfunctions and spectral functions}
From (\ref{V1explicit}) and (\ref{V2explicit}) we infer that the matrices $V_1$ and $V_2$ in the Lax pair
$$
\begin{cases}
	& \mu_x + i\zeta^2 [\sigma_3, \mu] = V_1\mu, \\
	& \mu_t + i\eta^2 [\sigma_3, \mu] = V_2\mu,
\end{cases}$$
are given explicitly by
\begin{align*}
& V_1 = \left(
\begin{array}{ll}
 -2 i \text{sech}(2 x) & -\frac{2 \sqrt{2} e^{2 i t+x} \zeta }{-i+e^{2
   x}} \\
 -\frac{2 \sqrt{2} e^{x-2 i t} \zeta }{i+e^{2 x}} & 2 i \text{sech}(2 x)
\end{array}
\right),
	\\
& V_2 = \left(
\begin{array}{ll}
 -2 i \text{sech}(2 x) & \frac{e^{2 i t-x} \left(2 \zeta ^2+e^{2 x}
   \left(2 i \zeta ^2+1\right)+i\right)}{\sqrt{2} \zeta  (-2 i \cosh (x)
   \sinh (x)-1)} \\
 -\frac{\sqrt{2} e^{x-2 i t} \left(2 i \zeta ^2+e^{2 x} \left(2 \zeta
   ^2+i\right)+1\right)}{\left(i+e^{2 x}\right)^2 \zeta } & 2 i
   \text{sech}(2 x)
\end{array}
\right).
\end{align*}
Moreover, as described in section \ref{solitonsec}, we compute the eigenfunction $\mu_2^{line}$ by the inverse scattering method on the line and use it to find the eigenfunction $\mu_3$ on the half-line.\footnote{Although the analysis of section \ref{solitonsec} was applied only to the initial data $u_0(x)$, it can be applied to $u(x,t)$ for any fixed time $t$ with obvious modifications.} We get
$$\mu_3(x,t,\zeta) = \left(
\begin{array}{ll}
 1+\frac{2}{\left(-i+e^{2 x}\right) \left(2 \zeta ^2-i\right)} & \frac{2
   \sqrt{2} e^{2 i t+x} \zeta }{\left(-i+e^{2 x}\right) \left(1-2 i \zeta
   ^2\right)} \\
 \frac{2 \sqrt{2} e^{ x-2 i t} \zeta }{\left(i+e^{2 x}\right) \left(2 i
   \zeta ^2+1\right)} & 1+\frac{2}{\left(i+e^{2 x}\right) \left(2 \zeta
   ^2+i\right)}
\end{array}
\right).$$
In particular,
$$\mu_3(x,t,\zeta) = \left(
\begin{array}{cc}
 1-\frac{2}{1+i e^{2 x}} & 0 \\
 0 & 1+\frac{2}{-1+i e^{2 x}}
\end{array}
\right) + O(\zeta), \qquad \zeta \to 0,$$
which in view of (\ref{Deltaints}) is seen to agree with (\ref{mu3zeta0}).
On the other hand, due to the simple $t$-dependence of $V_2$, it is possible to solve the $t$-part explicitly. Note that we must choose $T  < \infty$ when defining $\mu_1$, since $u^p(x,t)$ does not vanish as $t \to \infty$.
We obtain, for $x = 0$,\footnote{Here and in several other places we only give the second column of the $2 \times 2$ matrix since the first column can be obtained by symmetry.}
$$[\mu_1(0,t,\zeta)]_2 = \left(
\begin{array}{l}
 \frac{2 e^{3i\pi/4} e^{-2 i t \eta ^2} \left(e^{2 i t \left(\eta ^2+1\right)}-e^{2 i T \left(\eta
   ^2+1\right)}\right) \zeta  \left(\eta ^2+2\right)}{\left(2 \zeta ^2+1\right) \left(\eta ^2+1\right)} \\
 \frac{\eta ^2-e^{-2 i (t-T) \left(\eta ^2+1\right)}+2}{\eta ^2+1}
\end{array}
\right), \qquad 0 \leq t \leq T.$$
Using that $\eta = \zeta - \frac{1}{2\zeta}$ we infer that $[\mu_1(0,t,\zeta)]_2$ is bounded for $\zeta$ in $D_1 \cup D_3$ (recall that $\hbox{Im}\, \eta^2 > 0$ for $\zeta \in D_1 \cup D_3$) in accordance with equation (\ref{mu1largerdomain}). 
We also note that $[\mu_1(0,t,\zeta)]_2 \to (0, 1)^T$ as $\zeta \to 0$ in $D_3$ in accordance with (\ref{mu1zeta0}).

From these expressions for $\mu_1$ and $\mu_3$, we find 
\begin{align} \label{examplebaexplicit}
& \begin{pmatrix}
b(\zeta) \\ a(\zeta) \end{pmatrix} = [\mu_3(0,0,\zeta)]_2 = \left(
\begin{array}{l}
 -\frac{(1-i) \sqrt{2} \zeta }{2 \zeta ^2+i} \\
 1+\frac{1-i}{2 \zeta ^2+i}
\end{array}
\right),
	\\
& \begin{pmatrix}
B(\zeta) \\ A(\zeta) \end{pmatrix} = [\mu_1(0,0,\zeta)]_2 = \left(
\begin{array}{l}
 -\frac{2 e^{3i\pi/4} \left(-1+e^{2 i T \left(\eta ^2+1\right)}\right) \zeta  \left(\eta ^2+2\right)}{\left(2 \zeta
   ^2+1\right) \left(\eta ^2+1\right)} \\
 \frac{\eta ^2-e^{2 i T \left(\eta ^2+1\right)}+2}{\eta ^2+1}
\end{array}
\right).
\end{align}
$A(\zeta)$ and $B(\zeta)$ are bounded functions of $\zeta$ in $D_1 \cup D_3$ in accordance with the general theory. Moreover,
$$a(\zeta) \to 1, \quad b(\zeta) \to 0, \quad A(\zeta) \to 1, \quad B(\zeta) \to 0, \qquad \zeta \to \infty, \quad \zeta \in D_1.$$ 
As $\zeta \to 0$ in $D_2$, we find $a(\zeta) = -i + O(\zeta)$, $b(\zeta) = O(\zeta)$, while as $\zeta \to 0$ in $D_3$, $A(\zeta) = 1 + O(\zeta)$ and $B(\zeta) = O(\zeta)$. Using that $e^{-i\int_{(0,0)}^{(\infty, 0)}\Delta} = -i$ and $e^{-i\int_{(0,0)}^{(0,T)}\Delta} = 0$, we see that all these limits are as predicted by Propositions \ref{abprop} and \ref{ABprop}.

Since we know $\mu_3(x,t,\zeta)$ for all values of $x$ and $t$, as well as the spectral functions $s(\zeta)$ and $S(\zeta)$, we may use (\ref{seq}) and (\ref{Seq}) to determine the eigenfunctions $\mu_2(x,t,\zeta)$ and $\mu_1(x,t,\zeta)$ according to
$$\mu_2 = \mu_3 e^{-i(\zeta^2 x + \eta^2 t)\hat{\sigma}_3} s(\zeta)^{-1}, \qquad \mu_1 = \mu_2 e^{-i(\zeta^2 x + \eta^2 t)\hat{\sigma}_3} S(\zeta).$$
The expression for $\mu_2$ yields
\begin{align*} \label{examplemu2def}
&[\mu_2(x,t,\zeta)]_2 =
	\\
& \left(
\begin{array}{l}
 \frac{(1+i) \sqrt{2} e^{-2 i \left(x \zeta ^2+t \eta ^2\right)} \zeta  \left(-2 \zeta ^2+e^x \left(e^x \left(-2 i
   \zeta ^2-1\right)+(1+i) e^{2 i \left(x \zeta ^2+t \eta ^2+t\right)} \left(2 \zeta
   ^2+1\right)\right)-i\right)}{\left(-i+e^{2 x}\right) \left(4 \zeta ^4+1\right)} \\
 \frac{\left(2 \zeta ^2+1\right) \left(2 i \zeta ^2+e^{2 x} \left(2 \zeta ^2+i\right)+1\right)-(4+4 i) e^{-2 i x
   \zeta ^2+x-2 i t \left(\eta ^2+1\right)} \zeta ^2}{\left(i+e^{2 x}\right) \left(4 \zeta ^4+1\right)}
\end{array}
\right);
\end{align*}
the expression for $\mu_1$ is more complicated and will not be presented.
However, it can be explicitly verified using Mathematica that the expressions for the eigenfunctions $\mu_1, \mu_2, \mu_3$ derived in this subsection satisfy both the $x$- and $t$-parts of the Lax pair and the correct initial conditions.

\subsection{The jump matrices}
Now that we have computed the eigenfunctions, we may use (\ref{ddef}) and (\ref{Gammadef}) to compute $d(\zeta)$ and $\Gamma(\zeta)$. The result is
\begin{eqnarray} \label{exampledexplicit}
  & d(\zeta) = \frac{2 \zeta ^2+1}{2 \zeta ^2+i},
  	\\
  & \Gamma(\zeta) = -\frac{2 e^{i\pi/4} e^{-2 i T \left(\eta ^2+1\right)} \left(-1+e^{2 i T \left(\eta ^2+1\right)}\right) \zeta  \left(2 \zeta ^2+i\right)^2 \left(\eta
   ^2+2\right)}{\left(2 \zeta ^2+1\right)^3 \left(\eta ^2+1\right)}.
\end{eqnarray}  
Explicit expressions for the jump matrices are obtained from (\ref{J123def}). We find
\begin{eqnarray} \nonumber
J_1 = \left(
\begin{array}{ll}
 1 & 0 \\
 -\frac{2 e^{i\pi/4} e^{2 i \left(x \zeta ^2+t \eta ^2-T \left(\eta ^2+1\right)\right)} \left(-1+e^{2 i T \left(\eta ^2+1\right)}\right) \zeta  \left(2 \zeta
   ^2+i\right)^2 \left(\eta ^2+2\right)}{\left(2 \zeta ^2+1\right)^3 \left(\eta ^2+1\right)} & 1
\end{array}
\right),
	\\ \label{exampleJ123}
 J_3 =\left(
\begin{array}{ll}
 1 & \frac{2 e^{3i\pi/4} e^{-2 i \left(x \zeta ^2+t \eta ^2\right)} \left(-1+e^{2 i T \left(\eta ^2+1\right)}\right) \zeta  \left(i-2 \zeta ^2\right)^2 \left(\eta
   ^2+2\right)}{\left(2 \zeta ^2+1\right)^3 \left(\eta ^2+1\right)} \\
 0 & 1
\end{array}
\right),
	\\ \nonumber
J_4 = \left(
\begin{array}{ll}
 1 & \frac{(1-i) \sqrt{2} e^{-2 i \left(x \zeta ^2+t \eta ^2\right)} \zeta  \left(2 \zeta ^2-i\right)}{4 \zeta ^4+(2+2 i) \zeta ^2+i} \\
 \frac{(1-i) \sqrt{2} e^{2 i \left(x \zeta ^2+t \eta ^2\right)} \zeta  \left(2 \zeta ^2+i\right)}{4 i \zeta ^4+(2+2 i) \zeta ^2+1} & \frac{4 \zeta ^4+1}{\left(2 \zeta
   ^2+1\right)^2}
\end{array}
\right).
\end{eqnarray}
From these expressions $J_2$ is determined by $J_2 = J_3 J_4^{-1} J_1$.

\subsection{The RH problem}
The solution $M$ of the RH problem can be constructed from the formulas for the eigenfunctions by (\ref{MplusMminusdef}). It can be verified that this $M$ satisfies the jump conditions
$$M_2 = M_1 J_1, \quad M_2 = M_3 J_2, \quad M_4 = M_3 J_3, \quad M_4 = M_1 J_4,$$
with $J_i$, $i = 1, \dots,4$, given by (\ref{exampleJ123}).
In fact, these relations are identically satisfied for all values of $\zeta$---it is not necessary to restrict $\zeta$ to the appropriate contour since all functions have analytic continuations.

We would also like to verify the residue conditions (\ref{residue1})-(\ref{residue4}). This requires some care since we earlier assumed that none of the zeros of $a(\zeta)$ coincides with a zero of $d(\zeta)$. 
However, in the present case we find from (\ref{examplebaexplicit}) and (\ref{exampledexplicit}) that the zeros $\zeta_1, \zeta_2$ of $a(\zeta)$ coincide with the zeros $\lambda_1, \lambda_2$ of $d(\zeta)$. More precisely,
\begin{equation}\label{zeta12lambda12def}  
  \zeta_1 = \lambda_1 = \frac{i}{\sqrt{2}}, \quad \zeta_2 = \lambda_2 = -\frac{i}{\sqrt{2}}.
\end{equation}

We also previously assumed that the zeros of $a(\zeta)$ and $d(\zeta)$ were located away from the contour of the RH problem. But the zeros in (\ref{zeta12lambda12def}) lie on the contour. This difficulty can be overcome by assuming that $\zeta_1$ and $\zeta_2 = - \zeta_1$ have approached the contour from $D_1$ (so that $\bar{\zeta}_1$ and $\bar{\zeta}_2$ have approached the contour from $D_4$), and that $\lambda_1$ and $\lambda_2 = - \lambda_1$ have approached the contour from $D_2$ (so that $\bar{\lambda}_1$ and $\bar{\lambda}_2$ have approached the contour from $D_3$). Therefore we expect that the residue conditions (\ref{residue1})-(\ref{residue4}) at $\zeta_j$, $\bar{\zeta}_j$, $\lambda_j$, $\bar{\lambda}_j$ are satisfied by $M_1$, $M_4$, $M_2$, $M_3$, respectively. 

However, since the zeros of $a$ and $d$ coincide, the derivation of (\ref{residue1})-(\ref{residue4}) has to be reconsidered. Actually, the residue conditions (\ref{residue1})-(\ref{residue2}) remain the same as before: it can be verified that $M_1$ satisfies the residue condition (\ref{residue1}) at $\zeta_1$ and $\zeta_2$, whereas $M_4$ satisfies the residue condition (\ref{residue2}) at $\bar{\zeta}_1$ and $\bar{\zeta}_2$.
On the other hand, the conditions (\ref{residue3})-(\ref{residue4}) have to be modified as follows. Recall that the derivation of equation (\ref{residue3}) uses the first column of the equation $M_2 = M_1J_1$, namely the equation
$$[M_2]_1 = [M_1]_1 + \Gamma  e^{2i\theta} [M_1]_2.$$
The condition (\ref{residue3}) was obtained by evaluating the residue of this equation at $\zeta = \lambda_j$. Indeed, the residue of $[M_1]_1 = \mu_2^{(1)}/a$ at $\lambda_j$ vanishes if $a(\zeta)$ does not have a zero at $\lambda_j$. However, if $a(\lambda_j) = 0$ this residue gives a finite contribution and we need to amend the residue condition with this additional term. A similar argument applies to the residue at $\bar{\lambda}_j$. The correct conditions in the present case are therefore
\begin{align}\label{modifiedresidue3}
\underset{\lambda_j}{\text{Res}} [M_2(x,t,\zeta)]_1 = &
\underset{\lambda_j}{\text{Res}} [M_1(x,t,\zeta)]_1
+
\underset{\lambda_j}{\text{Res}} \, \Gamma(\zeta) \, e^{2i\theta(\lambda_j)} [M_2(x,t,\lambda_j)]_2, \quad j = 1, 2,	\\\label{modifiedresidue4}
\underset{\bar{\lambda}_j}{\text{Res}} [M_3(x,t,\zeta)]_2 =& 
\underset{\bar{\lambda}_j}{\text{Res}} [M_4(x,t,\zeta)]_2 
+
\underset{\bar{\lambda}_j}{\text{Res}} \, \overline{\Gamma(\bar{\zeta})} \,
e^{-2i\theta(\bar{\lambda}_j)} [M_3(x,t,\bar{\lambda}_j)]_1, \quad j = 1, 2.
\end{align}
These conditions can be directly verified for our solution $M$. This completes the verification of the RH problem.

\subsection{The reconstruction of $u$}
We can recover the solution $u(x,t)$ from the solution $M$ of the RH problem by following the steps introduced in (\ref{recoverqxt}). In particular,
$$u(x,t) = -\int_x^\infty 2im(x',t)e^{2i\int^{x'}_{0} \Delta} dx', \qquad m(x,t) = \lim_{\zeta \to \infty} \left(\zeta M(x,t, \zeta)\right)_{12},
$$
and
$$\Delta = 2 |m|^2 dx -2  \left(\int_x^\infty  \left(|m|^2\right)_t dx'\right)dt.$$
Using the expression for $M_1$ we find
$$m(x,t) = \lim_{\zeta \to \infty} (\zeta M_1(x,t,\zeta))_{12} = -\frac{\sqrt{2} e^{2 i t+x}}{1+i e^{2 x}}.$$
Hence
$$|m|^2 = \frac{2 e^{2 x}}{1+e^{4 x}},$$
and
$$\Delta = \frac{4 e^{2 x}}{1+e^{4 x}} dx.$$
Integration of
$$2im(x,t)e^{2i\int^{x}_{0} \Delta} = \frac{2 \sqrt{2} e^{2 i t+x} \left(-i+e^{2 x}\right)}{\left(i+e^{2 x}\right)^2},$$
yields
$$u(x,t) =-\frac{2 \sqrt{2} e^{2 i t+x}}{i+e^{2 x}},$$
which indeed is the solution $u^p(x,t)$ we started with in (\ref{simplestu}).

\section{An explicit example: the linearizable formalism}\label{examplelinearizablesec}\nequation
In the previous section we analyzed the IBV problem (\ref{exampleIBVproblem}) employing direct methods. However, since the boundary conditions are linearizable, we may alternatively study this problem by means of the linearizable scheme described in Theorem \ref{linearizableTh} and Lemma \ref{MtildeMlemma}. The linearizable approach amounts to constructing a modified RH problem with jump matrices $\tilde{J}_i$, $i = 1, \dots, 4$, constructed only in terms of the spectral function $a(\zeta)$ and $b(\zeta)$---the spectral functions $A(\zeta)$ and $B(\zeta)$ are eliminated by symmetry arguments. In this section we set up this modified RH problem for the IBV problem (\ref{exampleIBVproblem}) and verify that its solution $\tilde{M}(x,t,\zeta)$ is related to the original solution $M(x,t,\zeta)$ as predicted in (\ref{MtildeMrelations}). We will see explicitly that $\tilde{M}(x,t,\zeta)$ satisfies the jumps prescribed by the $\tilde{J}_i$'s, as well as the residue conditions of the modified RH problem. 

Using the expressions for $a(\zeta)$ and $b(\zeta)$ obtained in (\ref{examplebaexplicit}), we can compute the functions $\mathcal{D}(\zeta)$, $n(\zeta)$, $\tilde{\Gamma}(\zeta)$ as defined in the statement of Theorem \ref{linearizableTh}. The outcome is
\begin{equation}\label{nDGammatilde}
  n(\zeta) = 1, \qquad \mathcal{D}(\zeta) = \frac{2 i \zeta ^2+1}{2 \zeta ^2+i}, \qquad \tilde{\Gamma}(\zeta) = -\frac{2 e^{i\pi/4} \zeta  \left(2 \zeta ^2+i\right)}{4 \zeta ^4+(2-2 i) \zeta ^2-i}.
\end{equation}
Replacing $\Gamma$ by $\tilde{\Gamma}$ in the definition (\ref{J123def}) of the jump matrices, we obtain
\begin{align*}
&\tilde{J}_1 = \left(
\begin{array}{ll}
 1 & 0 \\
 -\frac{2 e^{i\pi/4} e^{2 i \left(x \zeta ^2+t \eta ^2\right)} \zeta  \left(2 \zeta ^2+i\right)}{4 \zeta ^4+(2-2 i) \zeta ^2-i} & 1
\end{array}
\right),
	\\
&\tilde{J}_2 =\left(
\begin{array}{ll}
 1 & 0 \\
 0 & 1
\end{array}
\right),
	\\
&\tilde{J}_3 = \left(
\begin{array}{ll}
 1 & -\frac{2 e^{3i\pi/4} e^{-2 i \left(x \zeta ^2+t \eta ^2\right)} \zeta  \left(2 \zeta ^2-i\right)}{4 \zeta ^4+(2+2 i) \zeta ^2+i} \\
 0 & 1
\end{array}
\right),
	\\
&\tilde{J}_4 =  \left(
\begin{array}{ll}
 1 & \frac{(1+i) \sqrt{2} e^{-2 i \left(x \zeta ^2+t \eta ^2\right)} \zeta  \left(2 \zeta ^2-i\right)}{4 i \zeta ^4-(2-2 i) \zeta ^2-1} \\
 -\frac{(1+i) \sqrt{2} e^{2 i \left(x \zeta ^2+t \eta ^2\right)} \zeta  \left(2 \zeta ^2+i\right)}{4 \zeta ^4+(2-2 i) \zeta ^2-i} & \frac{4 \zeta ^4+1}{\left(2 \zeta
   ^2+1\right)^2}
\end{array}
\right).
\end{align*}
It can be checked that $\tilde{J}_4 = J_4$ as expected.
Moreover, if we define $\tilde{M}_j$, $j = 1, \dots, 4$, according to (\ref{MtildeMrelations}), it can be verified that the jump conditions
$$\tilde{M}_2 = \tilde{M}_1 \tilde{J}_1, \quad \tilde{M}_2 = \tilde{M}_3 \tilde{J}_2, \quad \tilde{M}_4 = \tilde{M}_3 \tilde{J}_3, \quad \tilde{M}_4 = \tilde{M}_1 \tilde{J}_4,$$
are fulfilled. It can also be checked directly that $\tilde{M}$ satisfies the residue conditions (\ref{residue1})-(\ref{residue2}) and (\ref{modifiedresidue3})-(\ref{modifiedresidue4}) with $\Gamma$ replaced by $\tilde{\Gamma}$.

\subsection{The zeros of $\mathcal{D}(\zeta)$}
An important advantage of the methodology of \cite{F1997} is that it yields precise information about the long time asymptotic of the solution. By employing the Deift-Zhou formalism \cite{D-Z}, it can be shown that the solution of (\ref{GNLS2}) on the half-line will split for large $t$ into a collection of solitons traveling at constant speeds of order $1$, and away from these solitons the asymptotics will display a dispersive character (see \cite{F-I-S} for a precise description of the asymptotics of the solution to the half-line problem in the case of the NLS equation). The asymptotic solitons are generated by the zeros of $d(\zeta)$, which in the linearizable case coincide with the zeros of $\mathcal{D}(\zeta)$. As noted earlier, it is not possible in the present example to define $d(\zeta)$ when $T = \infty$, because $u(x,t)$ does not vanish as $t \to \infty$. Nevertheless, it is possible to find from (\ref{nDGammatilde}) the zeros of $\mathcal{D}(\zeta)$; they are located at $\zeta = \pm (1 + i)/2$. In view of our choice of parameters ($\Delta_0 = 1/\sqrt{2}$ and $\gamma = \pi/2$), we can verify from (\ref{parameterrelation}) that these zeros coincide with  the zeros corresponding to the one-soliton, which are located at 
$$\zeta_1^{line} = \frac{1 + i}{2} \quad \hbox{and} \quad  \zeta_2^{line}  = -\frac{1 + i}{2}.$$ 
Hence, as expected, the asymptotic soliton-content of the solution is correctly accounted for by the zeros of $\mathcal{D}(\zeta)$.

\appendix
\section{Behavior as $\zeta \to 0$} \label{appendixA}
\renewcommand{\theequation}{A.\arabic{equation}}\nequation
In this appendix we analyze the behavior of the eigenfunctions and the spectral functions near $\zeta = 0$. It is convenient to introduce the following notation:
$$\mu_{D_2} = \left(\mu_1^{(2)}, \mu_3^{(12)}\right), \qquad
\mu_{D_3} = \left(\mu_3^{(34)}, \mu_1^{(3)}\right).$$
Since $[\sigma_3, \mu] = 2\sigma_3\mu^{(o)}$, where $\mu^{(o)}$ denotes the off-diagonal part of $\mu$, the two columns of the Lax pair equations (\ref{mulax}) are independent of each other. Thus $\mu_{D_2}$ and $\mu_{D_3}$ are solutions of (\ref{mulax}) which are bounded and analytic in $D_2$ and $D_3$, respectively.
Substituting the expansion
\begin{equation}\label{linearmuexpansion}  
  \mu_{D_2} = D + Q_1 \zeta + Q_2 \zeta^2 + O\left(\zeta^3\right), \qquad \zeta \to 0, \quad \zeta \in D_2,
\end{equation}
where $D, Q_1, Q_2$ are independent of $\zeta$, into the $t$-part of (\ref{mulax}) it follows from the $O(1/\zeta^2)$-terms that $D$ is a diagonal matrix. Furthermore, one finds the following equations for the $O(1/\zeta)$ and the diagonal part of the $O(1)$ terms
\begin{equation}\label{Ozetaterms}  
  O(1/\zeta): \frac{i}{4}[\sigma_3, Q_1] = \frac{i}{2} \sigma_3 e^{-i\int_{(0,0)}^{(x,t)} \Delta \hat{\sigma}_3} U D \quad \text{i.e.} \quad Q_1^{(o)} = e^{-i\int_{(0,0)}^{(x,t)} \Delta \hat{\sigma}_3}UD,
\end{equation}
and
$$O(1): D_t = -\frac{i}{2}\sigma_3 U_x^2 D + \frac{i}{2} \sigma_3 \left(e^{-i\int_{(0,0)}^{(x,t)} \Delta \hat{\sigma}_3}U\right) Q_1^{(o)} \quad \text{i.e.} \quad D_t = -\frac{i}{2}\sigma_3(U_x^2 - U^2)D.$$
Similarly, substituting the expansion (\ref{linearmuexpansion}) into the $x$-part of (\ref{mulax}) it follows from the $O(1)$ terms that $D_x = -\frac{i}{2}\sigma_3U_x^2D$. 
Thus 
$$\mu_{D_2}(x,t,\zeta) = e^{i\int_{(x,t)}^{(\infty, 0)} \Delta \sigma_3}D_0 + O(\zeta), \qquad  \zeta \to 0, \quad \zeta \in D_2,$$
where $D_0$ is a constant matrix to be determined by the initial conditions of $\mu_1$ and $\mu_3$ at $(0,T)$ and $(\infty, t)$, respectively. 
An identical argument shows that the asymptotics of $\mu_{D_3}$ as $\zeta \to 0$ in $D_3$ are of the same form.
This yields the following asymptotics as $\zeta \to 0$:
\begin{subequations}\label{muzeta0}
\begin{align}\label{mu1zeta0}
  \mu_1(x,t,\zeta) = e^{i\int_{(x,t)}^{(0, T)} \Delta \sigma_3} + O(\zeta), \qquad \zeta \to 0,
	\\ \label{mu3zeta0}
  \mu_3(x,t,\zeta) = e^{i\int_{(x,t)}^{(\infty, 0)} \Delta \sigma_3} + O(\zeta), \qquad \zeta \to 0,
\end{align}
\end{subequations}
where the first and second columns of (\ref{mu1zeta0}) are valid for $\zeta$ in $D_2$ and $D_3$, respectively, while the first and second columns of (\ref{mu3zeta0}) are valid for $\zeta$ in $D_3$ and $D_2$, respectively.
From these asymptotics and the relations (\ref{Ssdef}) we deduce that the spectral functions $a(\zeta)$, $b(\zeta)$, $A(\zeta)$, $B(\zeta)$, have the following behavior as $\zeta \to 0$:
\begin{align}\label{baBAzeta0}
  \begin{pmatrix} b(\zeta) \\ a(\zeta) \end{pmatrix} 
  = [\mu_3(0,0,\zeta)]_2 
  = \begin{pmatrix} 0 \\ e^{-i\int_{(0,0)}^{(\infty, 0)}\Delta} \end{pmatrix} + O(\zeta), \qquad \zeta \to 0, \quad \zeta \in \bar{D}_2,
	\\ \nonumber
   \begin{pmatrix} B(\zeta) \\ A(\zeta) \end{pmatrix} 
  = [\mu_1(0,0,\zeta)]_2 
  = \begin{pmatrix} 0 \\ e^{-i\int_{(0,0)}^{(0, T)}\Delta} \end{pmatrix} + O(\zeta),  \qquad \zeta \to 0, \quad \zeta \in \bar{D}_3.
\end{align}
Finally, let us use these asymptotics to verify the jump condition (\ref{Mjump}) near $\zeta = 0$. It follows from the defintion (\ref{ddef}) of $d(\zeta)$ and (\ref{baBAzeta0}) that
\begin{equation}\label{dzeta0}  
  d(\zeta) = e^{i\int_{(\infty, 0)}^{(0,T)} \Delta} + O(\zeta), \qquad \zeta \to 0, \quad \zeta \in \bar{D}_2,
\end{equation}
and so, by (\ref{Jdef}) and (\ref{J123def}), the jump matrix $J_2$ has the following asymptotics as $\zeta \to 0$:
$$J_2 = I + O(\zeta), \qquad \zeta \to 0,\quad \zeta \in \bar{D}_2.$$
The jump condition (\ref{Mjump}) near $\zeta = 0$ is therefore
$$\left(\frac{\mu_1^{(2)}}{d(\zeta)}, \mu_3^{(12)}\right) = \left(\mu_3^{(34)}, \frac{\mu_1^{(3)}}{\overline{d(\bar{\zeta})}}\right) + O(\zeta), \qquad \zeta \to 0, \quad \zeta \in \bar{D}_2 \cap \bar{D}_3.$$
This equation can be easily verified using (\ref{muzeta0}) and (\ref{dzeta0}). 

\section{The linearized equation} \label{appendixB}
\renewcommand{\theequation}{B.\arabic{equation}}\nequation
As a general rule, before applying the methodology of \cite{F1997} to a nonlinear equation, it is illuminating to consider the linearized problem. Here we present the analysis of the linearized version of equation (\ref{GNLS2}) given by
\begin{equation}\label{gnlslinear}
  u_{tx} + u - 2iu_x - u_{xx} = 0.
\end{equation}
The equation adjoint to (\ref{gnlslinear}) is
\begin{equation}\label{adjoint}
  v_{tx} + v + 2iv_x - v_{xx} = 0.
\end{equation}
Multiplying (\ref{gnlslinear}) by $v_x$,  (\ref{adjoint}) by $u_x$, and adding the resulting equations, we find
\begin{equation}\label{divform}
  (u_x v_x)_t + (uv - u_xv_x)_x = 0.
\end{equation}
Equation (\ref{adjoint}) admits the solutions $v(x,t) = e^{2i(\zeta^2 x + \eta^2 t)}$, where $\zeta \in \C$ is a parameter and $\eta$ is defined in (\ref{etadef}). Substituting these solutions into (\ref{divform}), we deduce the existence of a scalar function $\mu(x,t,\zeta)$ such that
\begin{equation}\label{linearlaxdiff}
d(e^{2i(\zeta^2 x + \eta^2 t)} \mu(x,t,\zeta)) = W(x,t,\zeta),
\end{equation}
where the closed one-form $W(x,t,\zeta)$ is defined by
$$W = e^{2i(\zeta^2 x + \eta^2 t)} \left[2i\zeta^2 u_xdx + (2i\zeta^2 u_x - u)dt\right].$$
Equation (\ref{linearlaxdiff}) is a Lax pair for (\ref{gnlslinear}) in differential form; in components this Lax pair reads
\begin{equation}\label{linearpsilax}
\begin{cases}
	&  \mu_x + 2i\zeta^2 \mu = 2i \zeta^2 u_x, 	\\
	& \mu_t + 2i\eta^2 \mu = 2i\zeta^2u_x - u.
\end{cases}
\end{equation}
We define three solutions $\mu_j$, $j = 1,2,3$, of (\ref{linearlaxdiff}) by
\begin{equation}\label{linearmujdef}  
  \mu_j(x,t,\zeta) = \int_{(x_j, t_j)}^{(x,t)} e^{-2i(\zeta^2 x + \eta^2 t)}W(x',t',\zeta),
\end{equation}
where $(x_1, t_1) = (0, T)$, $(x_2, t_2) = (0, 0)$, and $(x_3, t_3) = (\infty, t)$. It follows that $\mu_1$, $\mu_2$, and $\mu_3$ are bounded and analytic in $D_3$, $D_4$, and $D_1\cup D_2$, respectively, where $\{D_j\}_1^4$ are as in (\ref{D1234def}). 
Substitution into (\ref{linearpsilax}) of the expansion
$$\mu = \mu_{(0)} + \frac{\mu_{(1)}}{\zeta} + \cdots, \qquad \zeta \to \infty,$$
shows that $\mu_2$ and $\mu_3$ have the following asymptotics as $\zeta \to \infty$:
$$\mu_2(x,t,\zeta) = u_x(x,t) + O(1/\zeta), \qquad \zeta \to \infty, \quad \zeta \in D_4,$$
$$\mu_3(x,t,\zeta) = u_x(x,t) + O(1/\zeta), \qquad \zeta \to \infty, \quad \zeta \in D_1.$$
Similarly, substitution into (\ref{linearpsilax}) of the expansion
$$\mu = \mu_{(0)} + \mu_{(1)} \zeta + \mu_{(2)} \zeta^2 + \cdots, \qquad \zeta \to 0,$$
shows that $\mu_1$ and $\mu_3$ have the following asymptotics as $\zeta \to 0$:
$$\mu_1(x,t,\zeta) = O(\zeta^2), \qquad \zeta \to 0, \quad \zeta \in D_3,$$
$$\mu_3(x,t,\zeta) = O(\zeta^2), \qquad \zeta \to 0, \quad \zeta \in D_2.$$
Thus, just like for the nonlinear problem, the limit of the eigenfunction as $\zeta \to \infty$ depends on the solution $u$, whereas the limit as $\zeta \to 0$ does not depend on $u$. In the case of the nonlinear problem we introduced a new eigenfunction (see (\ref{psimurelation})) in order to obtain an eigenfunction such that $\mu \to I$ as $\zeta \to \infty$. Here, it is easier to proceed without introducing a further transformation.

The eigenfunctions $\{\mu_j\}_1^3$ are related by
\begin{equation}\label{linearsmurelations}
\mu_3 - \mu_2 = e^{-2i(\zeta^2 x + \eta^2 t)} s(\zeta), \qquad
\mu_1 - \mu_2 = e^{-2i(\zeta^2 x + \eta^2 t)} S(\zeta),
\end{equation}
where the complex-valued spectral functions $s(\zeta)$ and $S(\zeta)$ are defined by
\begin{equation}\label{linearsSdef}
s(\zeta) := \mu_3(0,0,\zeta), \qquad 
S(\zeta) := \mu_1(0,0,\zeta).
\end{equation}
Introducing the notations
$$u_0(x) := u(x,0), \qquad g_0(t) := u(0, t), \qquad g_1(t) := u_x(0, t),$$
it follows from (\ref{linearmujdef}) and (\ref{linearsSdef}) that $s(\zeta)$  and $S(\zeta)$ admit the following expressions in terms of the initial and boundary values of $u$, respectively:
\begin{equation}\label{sSu0g01}
s(\zeta) = \int_\infty^0 e^{2i \zeta^2 x} 2i\zeta^2 u_{0x} dx, \qquad
S(\zeta) = \int_T^0 e^{2i\eta^2 t} (2i\zeta^2 g_1 - g_0)dt.
\end{equation}

We define the scalar function $M(x,t,\zeta)$ by
\begin{align}\label{linearMdef}
M_+ = \mu_3, \quad \zeta \in \bar{D}_1; \qquad
M_- = \mu_3, \quad \zeta \in \bar{D}_2;
		\\ \nonumber
M_+ = \mu_1, \quad \zeta \in \bar{D}_3; \qquad
M_- = \mu_2, \quad \zeta \in \bar{D}_4;
\end{align}
then (\ref{linearsmurelations}) implies that $M$ satisfies the jump condition
\begin{equation*}\label{linearMjump}  
  M_+(x,t,\zeta) - M_-(x,t, \zeta) = J(x,t,\zeta), \qquad \zeta \in \bar{D}_i \cap \bar{D}_j, \quad i,j = 1, \dots, 4,
\end{equation*}
where $J$ is defined by (\ref{Jdef}) and 
\begin{align}\label{linearJ1234}
& J_1 = 0, \qquad J_2 = e^{-2i(\zeta^2 x + \eta^2 t)}(S(\zeta) - s(\zeta)),
	\\ \nonumber
& J_3 = e^{-2i(\zeta^2 x + \eta^2 t)}S(\zeta),\qquad
J_4 = e^{-2i(\zeta^2 x + \eta^2 t)}s(\zeta).
\end{align}
Together with the normalization condition 
\begin{equation}\label{linearnormalization}  
  \lim_{\zeta \to 0} M(x,t,\zeta) = 0,
\end{equation}
this defines a scalar RH problem for $M$ with jump across the contour depicted in Figure \ref{GNLSRH2.pdf}. The solution of this RH problem is
\begin{equation}\label{linearRHsolution}
  M(x,t,\zeta) = \Lambda(x,t,\zeta) - \Lambda(x,t,0), 
\end{equation}
where
$$\Lambda(x,t,\zeta) 
= \frac{1}{2\pi i} \int_{\bar{D}_2 \cap \bar{D}_3} \frac{J_2(\zeta') d\zeta'}{\zeta' - \zeta}
  + \frac{1}{2\pi i} \int_{\bar{D}_3 \cap \bar{D}_4} \frac{J_3(\zeta') d\zeta'}{\zeta' - \zeta}
  + \frac{1}{2\pi i} \int_{\bar{D}_4 \cap \bar{D}_1} \frac{J_4(\zeta') d\zeta'}{\zeta' - \zeta}$$
and the contours $\bar{D}_i \cap \bar{D}_j$ are oriented as in Figure \ref{GNLSRH2.pdf}.

Equations (\ref{linearmujdef}) and (\ref{linearMdef}) express $M$ in terms of $u$ (the solution of the direct problem), whereas equations (\ref{linearJ1234}) and (\ref{linearRHsolution}) express $M$ in terms of the spectral functions $s(\zeta)$ and $S(\zeta)$ (the solution of the inverse problem). 
Taking the limit $\zeta \to \infty$ in (\ref{linearRHsolution}), we find $u_x(x,t) = -\Lambda(x,t,0)$. We can write $\Lambda(x,t,0)$ in terms of $s(\zeta)$ andÊ $S(\zeta)$ as
\begin{align*}\label{}
\Lambda(x,t,0)  = \frac{1}{2\pi i} \left(\int_{\partial D_1} + \int_{\partial D_2}\right) \frac{d\zeta}{\zeta}e^{-2i(\zeta^2 x + \eta^2 t)}s(\zeta)
+\frac{1}{2\pi i} \int_{\partial D_3} \frac{d\zeta}{\zeta} e^{-2i(\zeta^2 x + \eta^2 t)}S(\zeta),
\end{align*}
where $\partial D_j$ denotes the boundary of $D_j$ oriented so that $D_j$ lies to the left of $\partial D_j$.
Using the expressions (\ref{sSu0g01}) for $s(\zeta)$ and $S(\zeta)$ in this equation, we find the following integral representation of $u_x(x,t)$ in terms of the inital and boundary values:\footnote{Alternatively, an integral representation of $u(x,t)$ can be obtained by considering terms of $O(\zeta^2)$ in the limitÊ $\zeta \to 0$.}
\begin{align}\label{linearsolution}
u_x(x,t) = & \frac{1}{2\pi i} \left(\int_{\partial D_1} + \int_{\partial D_2}\right) \frac{d\zeta}{\zeta}\int_0^\infty dx' e^{2i \zeta^2 (x' - x) - 2i\eta^2 t} 2i\zeta^2 u_{0x}(x')
	\\ \nonumber
&+\frac{1}{2\pi i} \int_{\partial D_3} \frac{d\zeta}{\zeta}\int_0^T dt' e^{-2i\zeta^2x + 2i\eta^2 (t' - t)} (2i\zeta^2 g_1(t') - g_0(t')).
\end{align}

In order to verify that this formula gives the correct initial values, we note that if $t = 0$, then the second term on the right-hand side of (\ref{linearsolution}) involves the exponential $\exp[-2i\zeta^2x + 2i\eta^2 t']$ which is bounded and analytic in $D_3$. Thus, by deforming contours, we find that this term vanishes. On the other hand, setting $t=0$ in the first term on the right-hand side of (\ref{linearsolution}) and changing variables $k = 2\zeta^2$, $\frac{d\zeta}{\zeta} = \frac{dk}{2k}$, we find
$$u_x(x, 0) = \frac{1}{2\pi} \int_\R dk\int_0^\infty dx' e^{ik(x' - x)} u_{0x}(x')$$
The identity
\begin{equation}\label{deltafunction}
\int_\R dk e^{ik(x' - x)} = 2\pi \delta(x' - x)
\end{equation}
shows that indeed $u_x(x, 0) = u_{0x}(x)$.

In order to verify that the formula (\ref{linearsolution}) gives the correct boundary values, we note that if $x = 0$, then the first term on the right-hand side of (\ref{linearsolution}) involves the exponential $\exp[2i\zeta^2x' - 2i\eta^2 t]$ which is bounded and analytic in $D_2$. Thus, the integral along $\partial D_2$ vanishes. In order to simplify the integral along $\partial D_1$ we use the following global relation obtained by integrating the closed differential form $W$ around the rectangle in the $(x,t)$-plane with corners at $(\infty, 0), (0, 0), (0, T)$, and $(\infty, T)$:
\begin{align*}
\int_\infty^0 e^{2i\zeta^2 x'} 2i\zeta^2 u_{0x}(x') dx'
&+ \int_0^T e^{2i\eta^2t'}(2i\zeta^2 g_1(t') - g_0(t')) dt'
	\\
&+ \int_0^\infty e^{2i\zeta^2x' + 2i\eta^2T} 2i\zeta^2 u_x(x', T) dx' = 0.
\end{align*}
Using this in (\ref{linearsolution}) we find
\begin{align} \label{linearux0t}
u_x(0,t) =& \frac{1}{2\pi i} \left(\int_{\partial D_1} + \int_{\partial D_3}\right) \frac{d\zeta}{\zeta} \int_0^T e^{2i\eta^2(t' - t)}(2i\zeta^2 g_1(t') - g_0(t')) dt'
	\\ \nonumber
& +  \frac{1}{2\pi i} \int_{\partial D_1} \frac{d\zeta}{\zeta}\int_0^\infty e^{2i\zeta^2x' + 2i\eta^2(T - t)} 2i\zeta^2 u_x(x', T) dx'.
\end{align}
Now the second term on the right-hand side of (\ref{linearux0t}) vanishes since the integrand is bounded and analytic in $D_1$.
Moreover, performing the change of variables $k = 2\eta^2$ in the integrals along $\partial D_1$ and $\partial D_3$, we find
\begin{align*}
u_x(0,t) = &\frac{1}{2\pi i} \int_\R dk \left(\frac{2}{4 - \frac{1}{\zeta_+^4}} + \frac{2}{4 - \frac{1}{\zeta_-^4}}\right)
\int_0^T e^{ik(t' - t)} 2i g_1(t') dt'
	\\
& - \frac{1}{2\pi i} \int_\R dk \left(\frac{2}{4\zeta_+^2 - \frac{1}{\zeta_+^2}} + \frac{2}{4\zeta_-^2 - \frac{1}{\zeta_-^2}}\right) \int_0^T e^{ik(t' - t)} g_0(t') dt'
\end{align*}
where $\zeta_\pm$ satisfy\footnote{The choice of branch for the square root is immaterial, because all expressions are symmetric in $\zeta_+$ and $\zeta_-$.}
$$\zeta_\pm^2 = \frac{1}{4}(2 + k \pm \sqrt{4k + k^2}).$$
The identities 
$$\left(\frac{2}{4 - \frac{1}{\zeta_+^4}} + \frac{2}{4 - \frac{1}{\zeta_-^4}}\right) = \frac{1}{2},
\qquad
\left(\frac{2}{4\zeta_+^2 - \frac{1}{\zeta_+^2}} + \frac{2}{4\zeta_-^2 - \frac{1}{\zeta_-^2}}\right) = 0,$$
together with (\ref{deltafunction}) show that $u_x(0, t) = g_1(t)$ as expected.

These computations illustrate the following fact which is also valid for the nonlinear problem: although the eigenfunctions, in general, have essential singularities at $\zeta = 0$ and $\zeta = \infty$, the RH problem, which involves only the bounded, analytic pieces of the eigenfunctions, is perfectly regular near both these points. If a normalization condition is imposed at one of these points (for the nonlinear and linear problems, these conditions are given in equations (\ref{Mnormalization}) and (\ref{linearnormalization}), respectively), the asymptotics at the other point, which depend on the solution $u(x,t)$, are automatically of the required form.

 \bigskip
\noindent
{\bf Acknowledgement} {\it The authors are grateful to the referees for several helpful suggestions. Support is acknowledged from a Marie Curie Intra-European Fellowship and EPSRC.}

\bibliography{is}

\begin{thebibliography}{99}
\small

\bibitem{B-F-S}
A. Boutet de Monvel, A. S. Fokas, and D. Shepelsky, The analysis of the
global relation for the nonlinear Schr\"odinger equation on the half-line,
{\it Lett. Math. Phys.} {\bf 65} (2003), 199--212.

\bibitem{C-H}
R. Camassa and D. D. Holm, An integrable shallow water equation with
peaked solitons, {\it Phys. Rev. Lett.} {\bf 71} (1993),
1661--1664.

\bibitem{D-Z}
P. Deift and X. Zhou, A steepest descent method for oscillatory Riemann-
Hilbert problems. Asymptotics for the MKdV equation, 
{\it Ann. of Math.} {\bf 137} (1993), 295--368.

\bibitem{F}
A. S. Fokas, On a class of physically important integrable equations, {\it Phys.} D {\bf 87} (1995), 145--150.

\bibitem{F1997}
A. S. Fokas, A unified transform method for solving linear and certain nonlinear PDEs, 
{\it Proc. Roy. Soc. Lond.} A {\bf 453} (1997), 1411--1443.

\bibitem{F2002}
A. S. Fokas, Integrable nonlinear evolution equations on the half-line, 
{\it Comm. Math. Phys.} {\bf 230} (2002), 1--39.

\bibitem{F2004}
A. S. Fokas, Linearizable initial-boundary value problems for the sine-Gordon equation on the half-line, {\it Nonlinearity} {\bf 17} (2004), 1521--1534.

\bibitem{F2005}
A. S. Fokas, A generalised Dirichlet to Neumann map for certain nonlinear evolution PDEs, 
{\it Comm. Pure Appl. Math.} {\bf LVIII} (2005), 639--670.

\bibitem{F-I1992}
A. S. Fokas and A. R. Its, An initial-boundary value problem for the sine-Gordon equation in laboratory coordinates, 
{\it Theor. Math. Phys.} {\bf 92} (1992), 387--403.

\bibitem{F-I1994}
A. S. Fokas and A. R. Its, An initial-boundary value problem for the
Korteweg-de Vries equation, 
{\it Math. Comput. Simulation} {\bf 37} (1994), 293--321.

\bibitem{F-I1996}
A. S. Fokas and A. R. Its, The linearization of the initial-boundary value problem of the nonlinear Schr\"odinger equation,
{\it SIAM J. Math. Anal.} {\bf 27} (1996), 738--764. 

\bibitem{F-I-S}
A. S. Fokas, A. R. Its, and L.-Y. Sung, The nonlinear Schr\"odinger equation on the half-line, 
{\it Nonlinearity} {\bf 18} (2005), 1771--1822.

\bibitem{F-Liu}
A. S. Fokas and Q. M. Liu, Asymptotic integrability of water waves, 
{\it Phys. Rev. Lett.} {\bf 77} (1996), 2347--2351.

\bibitem{F-F}
B. Fuchssteiner and A. S. Fokas, Symplectic structures, their B{\"
a}cklund transformation and hereditary symmetries, {\it Physica D}
{\bf 4} (1981), 47--66.

\bibitem{L-F}
J. Lenells and A. S. Fokas, On a novel integrable generalization of the nonlinear Schr\"odinger equation,
{\it Nonlinearity} {\bf 22} (2009), 11--27.

\bibitem{LDNLS}
J. Lenells, The derivative nonlinear Schr\"odinger equation on the half-line, 
{\it Physica D} {\bf 237} (2008), 3008--3019.

\bibitem{Z-S1}
V. E. Zakharov and A. B. Shabat, A scheme for integrating the nonlinear equations of numerical physics by the method of the inverse scattering problem I, Funct. Anal. Appl. {\bf 8} (1974), 226--235.

\bibitem{Z-S2}
V. E. Zakharov and A. B. Shabat, A scheme for integrating the nonlinear equations of numerical physics by the method of the inverse scattering problem II, Funct. Anal. Appl. {\bf 13} (1979), 166--174.


\end{thebibliography}

\end{document}